\documentclass[aps,twocolumn,notitlepage,letterpaper]{revtex4-2}

\usepackage{graphicx}
\usepackage{float}
\usepackage{amsfonts}
\usepackage{amsmath}
\usepackage{amssymb}
\usepackage{bm}
\usepackage{mathrsfs}
\usepackage{graphicx}
\usepackage[caption=false]{subfig}
\usepackage{physics}
\usepackage{enumerate}
\usepackage{bm}
\usepackage{footmisc}

\usepackage[usenames,dvipsnames]{xcolor}
\usepackage[hyperindex,pdftex, breaklinks,colorlinks = true,linkcolor = blue,urlcolor=blue,citecolor=blue]{hyperref}

\newcommand{\bd}{\boldsymbol{\mathsf{d}}}
\newcommand{\bk}{\boldsymbol{k}}
\newcommand{\bG}{\boldsymbol{G}}
\newcommand{\bq}{\boldsymbol{\mathfrak{q}}}

\begin{document}
\title{Moir\'{e} Band Theory for M-Valley Twisted Transition Metal Dichalcogenides}

\author{Chao Lei}
\email{leichao.ph@gmail.com}

\author{Perry T.~Mahon}
\email{perry.mahon@austin.utexas.edu}

\author{A.~H.~MacDonald}
\affiliation{Department of Physics, University of Texas at Austin, Austin, Texas 78712, USA}

\thanks{C.~L.~and P.~T.~M.~contributed equally to this work.}

\begin{abstract}
We propose twisted bilayers of certain
group IV and IVB trigonal transition metal dichalcogenides (TMDs) MX$_{2}$ (M$=$Zr, Hf, Sn and X$=$S, Se) as
moir\'{e} materials.
In monolayer form these TMDs have conduction band minima near the three inequivalent Brillouin zone $M$ points and negligible spin-orbit coupling, 
implying six flavors of low-energy conduction band states.
The flavor sectors decouple at the single-particle level and 
in twisted bilayers are accurately described by emergent moir\'e-periodic Hamiltonians
that we derive from small-unit-cell density functional theory calculations.
Because the valley-projected Hamiltonians have large valley-dependent 
mass anisotropies and are 
time-reversal invariant, spontaneous valley polarization is signaled in transport by  
anisotropy instead of by the anomalous Hall and magnetic circular dichroism
signals commonly observed in graphene and $K$-valley TMD-based moir\'{e} multilayers.  
\end{abstract}

\maketitle

\textit{Introduction---} 
Moir\'{e} materials, artificial two-dimensional (2D) crystals with large lattice constants
created by forming moiré patterns from the crystal lattices of layered 2D materials, 
have emerged in recent years \cite{andrei2021marvels} as a rich
platform for experimental and theoretical discovery.
A 2D bilayer with relative twist angle ($\theta$) and/or lattice constant mismatch between layers is a moir\'{e} material only if its (electronic) states near the Fermi energy are described by a set of 
emergent Hamiltonians with the moir\'e periodicity \cite{bistritzer2011moire,MacDonald2014}.
Moir\'e materials support low-energy 
Bloch bands that can have a small bandwidth and are often topologically nontrivial.
Moreover, the large artificial lattice constants allow the number of 
electrons per moir\'{e} unit cell to be significantly tuned
with electrical gates, effectively moving through
an artificial periodic table without chemical doping.  
Materials of this type therefore serve as tunable quantum systems
in which strong-correlation and topological electron physics can be studied, often together. 

The moir\'e materials that have been studied experimentally to date are almost exclusively 
formed from either graphene or group VI transition metal dichalcogenide (TMD) 2D layers \cite{andrei2021marvels,mak2022semiconductor}.
Accordingly, there has been a focus on 
developing theoretical models to describe the low-energy states in these specific materials \cite{CastroNeto2007,bistritzer2011moire,Bernevig_TBG1,Senthil2018,wu2018hubbard,wu2019topological,devakul2021magic,angeli2021gamma}.
In a generic weakly coupled multilayer material, states near the Fermi energy are linear combinations of the near-Fermi-energy Bloch states of the constituent crystalline 2D layers.
In moir\'{e} materials, those Bloch states reside only in small, well-separated Brillouin zone (BZ) pockets, and Bloch states with crystal momenta in distinct pockets decouple; separate periodic models emerge, one associated with each low-energy point (i.e., valley) \cite{MacDonald2014}.
In graphene and group VI TMDs the low-energy points are usually the $K$ or $\Gamma$ points in their respective triangular lattice BZs. 
Most attention has been given to developing models for $K$-valley materials.
Since there are two inequivalent $K$ points related by time-reversal 
symmetry (TRS) in triangular lattices, each valley-projected moir\'e band Hamiltonian 
does not generically display TRS.
This property is the root cause   \cite{Randeria_2019,Dai2019,Xie_2020,wu2019topological,reddy2023fractional,tao2024valley} of the anomalous Hall effects and magnetic circular dichroism, 
and of the integer and fractional quantum anomalous Hall effects, seen \cite{Sharpe_2019,Serlin_2020,Li_2021,expt1,expt2,expt3,expt4,expt5} in these materials.

In this Letter, we derive moir\'e band models for
the low-energy conduction band states in twisted bilayers of certain group IV and IVB TMDs -- 
namely, 1T-MX$_{2}$ for M$=$Zr, Hf, Sn and X$=$S, Se.
We use homobilayer 1T-HfS$_{2}$ twisted relative to AA-stacking 
\footnote{By AA-stacking we mean that the two layers are stacked in a perfectly aligned configuration.} 
(denoted $t$HfS$_2$) as a representative example. 
In monolayer 1T-HfS$_2$, the conduction band minima (CBM) are at the three inequivalent $M$ points of its triangular lattice BZ, and spin-orbit coupling is negligible \cite{yan20182d,HfS2_DFT}.
Thus, there are six flavors of low-energy conduction band states that decouple at the single-particle level. 
In $t$HfS$_2$, we derive 
emergent periodic Hamiltonians for each sector
using
Wannier function (WF) processed density functional theory (DFT) data, which apply in twisted bilayers
with small $\theta$ and $n$-type electrostatic doping.  
Moir\'e band models for the other listed 1T-MX$_2$ materials take the same form and are 
defined by model parameters detailed in Supplemental Material \footnote{See Supplemental Material at [url], including details of the DFT calculations, interlayer distance, energy landscape, and interlayer and intralayer parameters landscape, as well as the Fourier components and moir\'e parameters of HfS$_2$, HfSe$_2$, ZrS$_2$, ZrSe$_2$, SnS$_2$, SnSe$_2$. The Supplemental Material also contains Refs.~\cite{QE2009,strain_hfs2,nam2017lattice}}.
In these $M$-valley materials, the TRS of the constituent layers is inherited by each valley-projected moir\'{e} band Hamiltonian, which is expected since the CBM occur at time-reversal invariant momenta (TRIM).
Therefore, unlike in $K$-valley systems, 
anomalous Hall physics and magnetic circular dichroism cannot be used to detect spontaneous valley polarization, which is expected when correlations are strong.
However, the strong CBM anisotropy of the constituent layers is also inherited in the moir\'e bands for each valley, especially in Hf- and Zr-based materials.
Thus, transport anisotropy should provide a clear experimental signal of valley polarization.

\begin{figure}[t!]
	\centering
	\includegraphics[width= 1.0\linewidth]{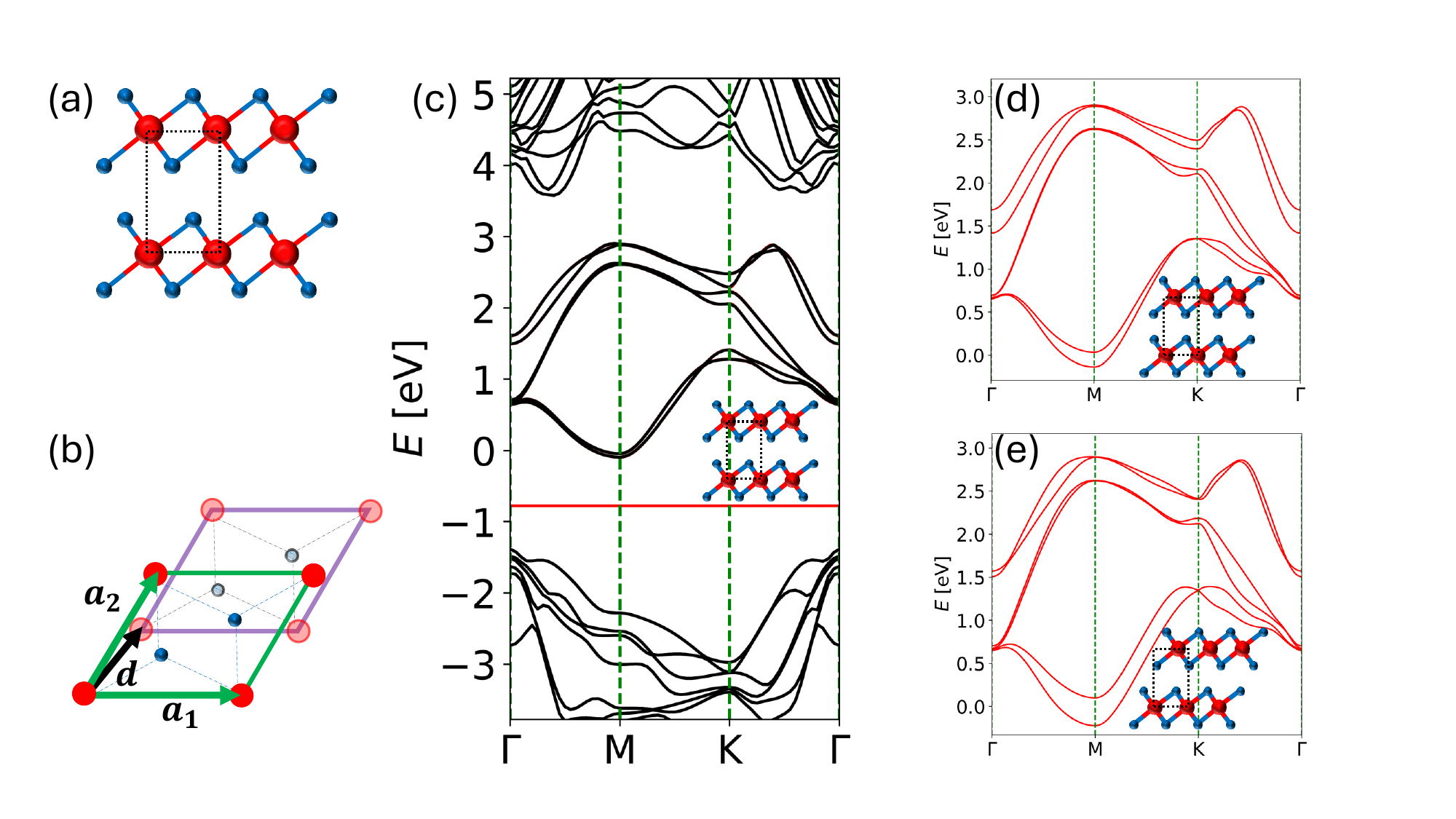}
	\caption{(a) Side view of (AA-stacked) crystalline bilayer 1T-HfS$_{2}$. (b) Top view of a bilayer with displacement vector $\bd$. Red (blue) dots identify the positions of Hf (S) ions. (c) Band structure for $\bd=\boldsymbol{0}$ obtained from DFT (black) and Wannier tight-binding model(red).
    (d) and (e) Low-energy conduction bands from Wannier basis for $\bd=(\boldsymbol{a}_{1}+\boldsymbol{a}_{2})/3$ and $\bd=2(\boldsymbol{a}_{1}+\boldsymbol{a}_{2})/3$.}
	\label{fig:Bands}
\end{figure}

\textit{Local electronic structure in bilayer 1T-HfS$_{2}$---}
The construction of a moir\'e band model via the local displacement approach (see Appendix for details) 
for a generic incommensurate bilayer employs a family of Bloch Hamiltonians $H(\bd)$
for (fictitious) crystalline bilayers -- constructed
as aligned stacks of the quasicrystal’s (adjusted) constituent layers, each having Bravais
lattice $\Lambda$ -- 
that is parametrized by an in-plane relative displacement $\bd\in\mathbb{R}^2$ between layers 
(see Fig.~\ref{fig:Bands}(b)) with respect to which $H(\bd)$ is $\Lambda$-periodic.
For each $\bd$, DFT calculations are performed, and bands in the energy range of interest are Wannierized to obtain Hamiltonian matrix elements,
\begin{align}
    \mathscr{H}^{l,\alpha_{l}}_{l',\beta_{l'}}(\boldsymbol{k};\boldsymbol{\mathsf{d}})
    \equiv\bra{\tilde{\psi}_{(l,\alpha_{l}),\boldsymbol{k}}(\boldsymbol{\mathsf{d}})}H(\boldsymbol{\mathsf{d}})\ket{\tilde{\psi}_{(l',\beta_{l'}),\boldsymbol{k}}(\boldsymbol{\mathsf{d}})}
\end{align}
that contain microscopic data needed to construct the moir\'{e} band model.
Here, $\big|\tilde{\psi}_{(l,\alpha_l),\boldsymbol{k}}(\boldsymbol{\mathsf{d}})\big\rangle$ are the Bloch-type vectors defined by the inverse Bloch-Floquet transform \cite{Marzari2012} of the energetically relevant WFs, where $l\in \{\mathfrak{b},\mathfrak{t}\}$ is a layer label (top $\mathfrak{t}$ and bottom $\mathfrak{b}$) and $\alpha_{l}$ an orbital-type label in layer $l$.
Due to the $\bd$-dependence of the WFs,
\begin{align}
	\bar{\mathscr{H}}^{l,\alpha_{l}}_{l',\beta_{l'}}(\boldsymbol{k};\boldsymbol{\mathsf{d}})\equiv
	e^{i(\delta_{l',\mathfrak{t}}-\delta_{l,\mathfrak{t}})\boldsymbol{k}\cdot\boldsymbol{\mathsf{d}}}\mathscr{H}^{l,\alpha_{l}}_{l',\beta_{l'}}(\boldsymbol{k};\boldsymbol{\mathsf{d}})
\end{align}
is $\Lambda$-periodic in $\bd$ and admits a Fourier series expansion with coefficients $\bar{\mathscr{H}}^{l,\alpha_{l}}_{l',\beta_{l'}}(\boldsymbol{k};\boldsymbol{G})$
for $\boldsymbol{G}\in\Lambda^{*}$.
Bar accents are used below to identify phase-modified quantities.
We now apply this approach, focusing on the low-energy conduction band states in $t$HfS$_2$.

Monolayer 1T-HfS$_{2}$ has a triangular lattice $\Lambda$ with Hf ions located at Bravais lattice sites around which S ions form octahedral cages \cite{yan20182d}.
Trigonal monolayers and AA-stacked
bilayers (with any $\bd$) have CBM near the three $M$ points ($\mathcal{K}\equiv\{\boldsymbol{M}_{1},\boldsymbol{M}_{2},\boldsymbol{M}_{3}\}$)
and a center of inversion symmetry
that leads to a Kramers spin-degeneracy at each $\bk\in\text{BZ}$.
For 1T-HfS$_{2}$ bilayers, we find that spin-orbit interactions
have negligible impact on the six spin-degenerate lowest energy conduction bands
\footnote{DFT band structure calculations for 1T-HfS$_{2}$ monolayers that employ different pseudopotentials can result in more significant, albeit still small, SOI in the low-energy conduction bands (see, e.g., Fig.~9 of Ref.~\cite{SOI_Buchner2018}). 
However, even there, near the $M$ valley SOI has almost no impact.},
and $s_z$ is approximately a good quantum number throughout the BZ.
These bands remain energetically isolated from all other bands throughout the BZ (as shown in Figs.~\ref{fig:Bands}(c)-(e))
and vary weakly with $\bd$ \footnotemark[2], 
demonstrating the weak interlayer 
coupling needed for the accuracy of local displacement moir\'e band models. 
Thus, we take $\alpha_{\mathfrak{b}}, \alpha_{\mathfrak{t}} =(\alpha,s_z)$ with $\alpha\in\{1,2,3\}$, and within each $s_z$ spin sector the lowest energy conduction bands are completely described throughout the 
BZ by a six-orbital Wannier tight-binding Hamiltonian 
$H_{\text{TB}}(\bd)$.

In principle, for the low-energy conduction states in $t$HfS$_2$, we could construct a moir\'e band model 
with three orbital times two spin degrees of freedom in each layer.
For each $\bd$, we find \footnotemark[2], however, that near each $\boldsymbol{M}_{\nu}\in\mathcal{K}$, the two lowest energy bands of $H_{\text{TB}}(\bd)$ are separated from the other four bands by $\approx 2$ eV (see Figs.~\ref{fig:Bands}(c)-(e)).  This motivates the construction of a 
simpler and more physically transparent model, with only layer and spin
degrees of freedom, that still
accounts for the higher energy states perturbatively
via L\"{o}wdin partitioning \cite{lowdin1951note}.
In each $s_z$ spin sector, we begin with the $6\times 6$ matrix representation of $H_{\text{TB}}(\bd)$ in the basis
$|\tilde{\psi}_{(l,\alpha,s_z),\bk}(\bd)\big\rangle$,
whose 
elements are $\mathscr{H}^{l,\alpha,s_z}_{l',\beta,s_z}(\boldsymbol{k};\boldsymbol{\mathsf{d}})$,
and re-express it
in the basis of Bloch energy eigenvectors at 
the relevant valley center $\boldsymbol{M}_{\nu}\in\mathcal{K}$ and
at the (arbitrary) reference displacement $\bd_{0}=\boldsymbol{0}$.
This yields a matrix of the form
\begin{align}
	\mathcal{H}_{\boldsymbol{M}_{\nu}}(\boldsymbol{k};\boldsymbol{\mathsf{d}})=
	\begin{pmatrix}
		\mathcal{H}_{\text{lowE}}(\boldsymbol{k};\boldsymbol{\mathsf{d}}) & \mathcal{T}(\boldsymbol{k};\boldsymbol{\mathsf{d}})\\
		\mathcal{T}^{\dagger}(\boldsymbol{k};\boldsymbol{\mathsf{d}})& \mathcal{H}_{\text{highE}}(\boldsymbol{k};\boldsymbol{\mathsf{d}})
	\end{pmatrix},
\end{align}
where $\mathcal{H}_{\text{lowE}}(\boldsymbol{k};\boldsymbol{\mathsf{d}})$ is a $2\times 2$ matrix, $\mathcal{H}_{\text{highE}}(\boldsymbol{k};\boldsymbol{\mathsf{d}})$ is a $4\times 4$ matrix, and $\mathcal{T}(\boldsymbol{k};\boldsymbol{\mathsf{d}})$ is a $2\times 4$ matrix.
Following Appendix B of Ref.~\cite{WinklerBook},
we identify $H^{0}\equiv {H}_{\text{lowE}}(\boldsymbol{M}_{\nu};\boldsymbol{\mathsf{d}}=\boldsymbol{0})$ and $H'\equiv {H}_{\boldsymbol{M}_{\nu}}(\boldsymbol{k};\boldsymbol{\mathsf{d}})-{H}_{\text{lowE}}(\boldsymbol{M}_{\nu};\boldsymbol{\mathsf{d}}=\boldsymbol{0})$ 
\footnote{Here we use the notation that ${H}(\boldsymbol{k};\boldsymbol{\mathsf{d}})$ is the operator in the electronic Hilbert space defined by the matrix representation $\mathcal{H}(\boldsymbol{k};\boldsymbol{\mathsf{d}})$ in the basis $\big(\ket{\psi_{n,\boldsymbol{M}_{\nu}}(\bd_{0})}\big)_{n\in\{1,\ldots,6\}}$.}.
Taking the zero of energy at the $\bd=\boldsymbol{0}$ CBM (i.e., the $H^0$ eigenvalues, which are approximately equal), the effective Hamiltonian in the basis of eigenvectors of $H^0$ is given to second order in $\mathcal{T}$ by the $2\times 2$ matrix
\begin{align}
    \mathcal{H}^{\boldsymbol{M}_{\nu}}_{\text{eff}}(\boldsymbol{k};\boldsymbol{\mathsf{d}})
    &=\mathcal{H}_{\text{lowE}}(\boldsymbol{k};\boldsymbol{\mathsf{d}})-
    \mathcal{T}^{\dagger}(\boldsymbol{k};\boldsymbol{\mathsf{d}})\mathcal{H}_{\text{highE}}^{-1}(\boldsymbol{k};\boldsymbol{\mathsf{d}}) \mathcal{T}(\boldsymbol{k};\boldsymbol{\mathsf{d}}).
    \label{Heff}
\end{align}

\begin{figure}[t!]
	\centering
	\includegraphics[width= 0.9\linewidth]{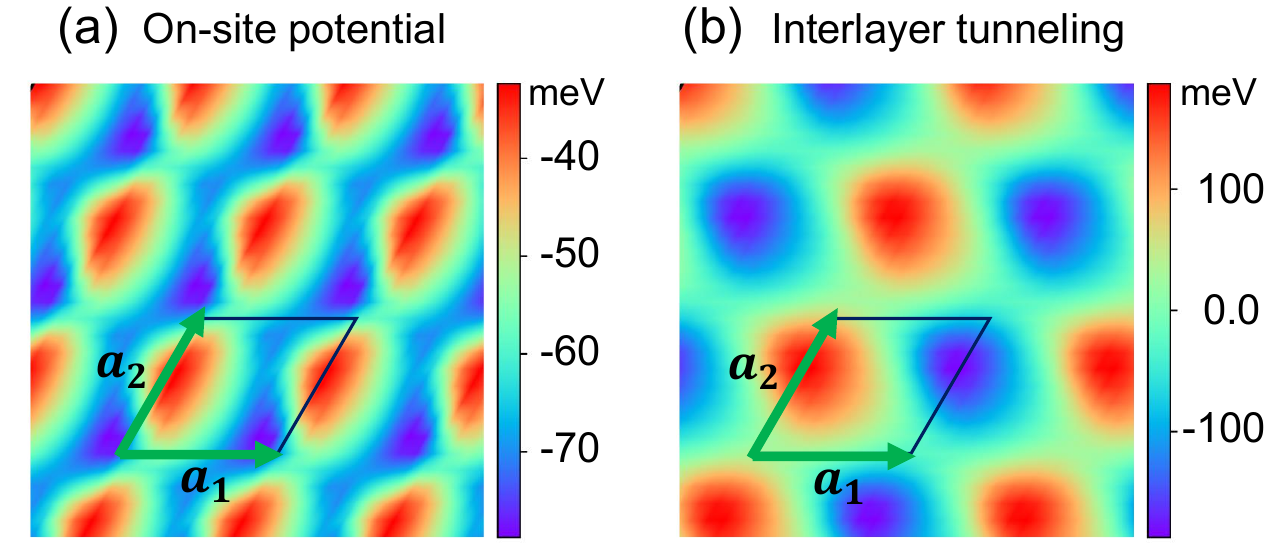}
	\caption{
    (a) Intralayer potential $\Delta_{l}(\boldsymbol{M}_{1};\bd)$ and 
    (b) interlayer tunneling amplitude $\Delta_{T}(\boldsymbol{M}_{1};\bd)$ 
    {\it vs.}~stacking $\bd$ in 1T-HfS$_2$ AA-stacked crystalline bilayers.  
    The parallelograms identify a unit cell of HfS$_2$.
    The interlayer tunneling is periodic in a doubled unit cell 
    (along $\boldsymbol{a}_{1}$ for valley $\boldsymbol{M}_{1}=\boldsymbol{b}_{1}/2$)
    due to the $\bd$-dependence of the WFs discussed in the main text. 
    The $\bd$-dependence of these terms
    is mapped to position-dependence on the moir\'{e} scale when modeling twisted bilayers.}
	\label{fig:realspace}
\end{figure}

Diagonalizing $\mathcal{H}^{\boldsymbol{M}_{\nu}}_{\text{eff}}(\boldsymbol{k};\boldsymbol{\mathsf{d}})$ for each $\bd$ and each $\bk$ near $\boldsymbol{M}_{\nu}$, we find that its eigenvectors $\big|\phi^{\boldsymbol{M}_{\nu}}_{\pm,\bk}(\bd)\big\rangle$ are even and odd combinations of layer-polarized Bloch-type states: $\big|\phi^{\boldsymbol{M}_{\nu}}_{\pm,\bk}(\bd)\big\rangle=\big|\tilde{\phi}^{\boldsymbol{M}_{\nu}}_{\mathfrak{b},\bk}(\bd)\big\rangle + S_{\boldsymbol{M}_{\nu}}(\bd)\big|\tilde{\phi}^{\boldsymbol{M}_{\nu}}_{\mathfrak{t},\bk}(\bd)\big\rangle$, where $\big|\tilde{\phi}^{\boldsymbol{M}_{\nu}}_{l,\bk}(\bd)\big\rangle=\sum_{\alpha=1}^{3}C^{\boldsymbol{M}_{\nu}}_{\alpha}(\bk,\bd)\big|\tilde{\psi}_{(l,\alpha,s_z),\boldsymbol{M}_{\nu}}(\bd_{0})\big\rangle$ and $S_{\boldsymbol{M}_{\nu}}(\bd)=\pm1$ depending on $\bd$ and $\boldsymbol{M}_{\nu}$. 
When transformed back to a layer representation,
the average of the eigenvalues can be interpreted as a potential that is 
identical in each layer 
and half the difference as a real tunneling amplitude between layers.
The sign of the tunneling is chosen for each $\bd$ such that the lowest energy CBM state $\big|\phi^{\boldsymbol{M}_{\nu}}_{-,\bk}(\bd)\big\rangle$ has layer parity consistent with that determined by DFT \footnotemark[2].
These properties are expected, since each $\boldsymbol{M}_{\nu}\in\mathcal{K}$ is a TRIM, and 
for each $\bd$, the crystalline bilayer has inversion symmetry; thus, the top and bottom intralayer potentials must be identical, and the interlayer tunneling amplitude must be real-valued.
The $\bd$-dependence of the intralayer potential $\Delta_{l}(\boldsymbol{M}_{\nu};\boldsymbol{\mathsf{d}})$ and of the interlayer tunneling $\Delta_{T}(\boldsymbol{M}_{\nu};\boldsymbol{\mathsf{d}})$ for valley $\boldsymbol{M}_{1}=\boldsymbol{b}_{1}/2$
is plotted in Fig.~\ref{fig:realspace}.
In Table \ref{table:FourierComponents}, we list all sizable Fourier components
of $\Delta_{l}(\boldsymbol{M}_{1};\boldsymbol{\mathsf{d}})$
and $\bar{\Delta}_{\mathfrak{t}}^{\mathfrak{b}}(\boldsymbol{M}_{1};\bd)\equiv e^{i\boldsymbol{M}_{1}\cdot\bd}\Delta_{T}(\boldsymbol{M}_{1};\bd)$ for top-to-bottom tunneling.  
As anticipated, only a small number of parameters are needed to define the 
moir\'e band model for valley $\boldsymbol{M}_{1}$ in $t$HfS$_2$. 
Similar analyses apply to the other identified 1T-MX$_2$ homobilayers, of which the model parameters are listed in Supplemental Material \footnotemark[2].
The valley-projected Hamiltonians for other valleys are 
related to that for valley $\boldsymbol{M}_{1}$ by the $C_{3}$ rotational symmetry of monolayer 1T-HfS$_2$, as we now describe. 

\textit{Symmetry analysis---} 
The spatial symmetries of (AA-stacked) 1T-HfS$_{2}$ crystalline bilayers (which are also present in the other candidate materials)
constrain the form of the matrix elements $\mathscr{H}^{l,\alpha,s_z}_{l',\beta,s_z}(\boldsymbol{k};\boldsymbol{\mathsf{d}})$, and therefore, also that of the effective masses $m_{a,l}^{\boldsymbol{M}_{\nu}}$ (defined below), $\Delta_{l}(\boldsymbol{M}_{\nu};\boldsymbol{\mathsf{d}})$, and $\Delta_{T}(\boldsymbol{M}_{\nu};\boldsymbol{\mathsf{d}})$.
In bulk, each Hf ion location in 1T-HfS$_2$ has $D_{3d}$ point group symmetry.
This point group symmetry is inherited in the bilayer
with $\bd=\boldsymbol{0}$ (albeit no longer about Hf sites) and, in fact, 
each group element has a counterpart in a bilayer with general $\bd$, namely:
(a) a center of inversion symmetry, which implies that the effective masses and the potentials
are layer independent ($m_{a,\mathfrak{b}}^{\boldsymbol{M}_{\nu}}=m_{a,\mathfrak{t}}^{\boldsymbol{M}_{\nu}}$ and $\Delta_{\mathfrak{b}}(\boldsymbol{M}_{\nu};\boldsymbol{\mathsf{d}})=\Delta_{\mathfrak{t}}(\boldsymbol{M}_{\nu};\boldsymbol{\mathsf{d}})$) and that the tunneling is real-valued
($\Delta_{T}(\boldsymbol{M}_{\nu};\boldsymbol{\mathsf{d}})=\Delta_{T}^{*}(\boldsymbol{M}_{\nu};\boldsymbol{\mathsf{d}})$).
Here and below, we use that $\boldsymbol{M}_{\nu}$ is equivalent to $-\boldsymbol{M}_{\nu}$;
(b) a symmetry that involves the rotation of the top layer by $\Theta=2\pi n/3$ for $n\in\mathbb{Z}$ about the surface-normal $\hat{\boldsymbol{e}}_{z}$ relative to the bottom layer,
which implies 
$m_{a,l}^{\boldsymbol{M}_{\nu}}=m_{a,l}^{R^{z}_{\Theta}(\boldsymbol{M}_{\nu})}$ 
\footnote{``$\parallel$'' and ``$\perp$'' on LHS and RHS are with respect to $\boldsymbol{M}_{\nu}$ and $R^{z}_{\Theta}(\boldsymbol{M}_{\nu})$, respectively.} 
and
$\Delta_{\mu}(\boldsymbol{M}_{\nu};\boldsymbol{\mathsf{d}})=
\Delta_{\mu}(R^{z}_{\Theta}(\boldsymbol{M}_{\nu});R^{z}_{\Theta}(\boldsymbol{\mathsf{d}}))$ for $\mu\in\{b,t,T\}$; and
(c) three symmetries that involve a total rotation by $\pi$ about in-plane axes $\hat{\boldsymbol{e}}_{i}$ 
($i\in\{\text{A,\,B,\,C}\}$, each of which is parallel to one of the three lines connecting nearest-neighbor Hf ions and positioned midway between the layers)
followed by $\bd\rightarrow -R^{i}_{\pi}(\bd)$, 
which for the $\boldsymbol{M}_{1}=\boldsymbol{b}_{1}/2$ valley implies $\Delta_{\mathfrak{b}}(\boldsymbol{M}_{1};\boldsymbol{b}_{2})=\Delta_{\mathfrak{t}}(\boldsymbol{M}_{1};-\boldsymbol{b}_{1}-\boldsymbol{b}_{2})$, 
$\bar{\Delta}_{\mathfrak{t}}^{\mathfrak{b}}(\boldsymbol{M}_{1};\boldsymbol{0})=\bar{\Delta}_{\mathfrak{t}}^{\mathfrak{b}}(\boldsymbol{M}_{1};-\boldsymbol{b}_{1})^{*}$, 
$\bar{\Delta}_{\mathfrak{t}}^{\mathfrak{b}}(\boldsymbol{M}_{1};\boldsymbol{b}_2)\in\mathbb{R}$, $\bar{\Delta}_{\mathfrak{t}}^{\mathfrak{b}}(\boldsymbol{M}_{1};-\boldsymbol{b}_{1}-\boldsymbol{b}_{2})\in\mathbb{R}$,
$\bar{\Delta}_{\mathfrak{t}}^{\mathfrak{b}}(\boldsymbol{M}_{1};\boldsymbol{b}_{1}+\boldsymbol{b}_{2})=\bar{\Delta}_{\mathfrak{t}}^{\mathfrak{b}}(\boldsymbol{M}_{1};-\boldsymbol{b}_{1}+\boldsymbol{b}_{2})^{*}$.
Since the $s_z$ spin sectors are identical, TRS implies $\bar{\Delta}_{\mathfrak{t}}^{\mathfrak{b}}(\boldsymbol{M}_{1};\boldsymbol{G})=\bar{\Delta}_{\mathfrak{t}}^{\mathfrak{b}}(\boldsymbol{M}_{1};-\boldsymbol{G}-\boldsymbol{b}_{1})^{*}$; thus, $\Delta_{T}(\boldsymbol{M}_{\nu};\boldsymbol{\mathsf{d}})\in\mathbb{R}$ as in (a).
This set of relations explains the most prominent structure of Fig.~\ref{fig:realspace} via Table~\ref{table:FourierComponents}.

\begin{table}[t!]
\begin{tabular}{||c c c||}
    \hline
    \, $\boldsymbol{G}=n_{1}\boldsymbol{b}_{1}+n_{2}\boldsymbol{b}_{2}$ \, & \, $l=l'$ ($10^{-3}$ eV) \, & \, $l\neq l'$ ($10^{-3}$ eV) \, \\ [0.5ex] 
    \hline\hline
    $n_{1}=0, n_{2}=0$ & arb. &  $\bm{42.0-8.0i}$  \\
    \hline\hline
    $n_{1}=1, n_{2}=0$ & $3.5+3.5i$   & $-0.30-1.59i$  \\
    \hline
    $n_{1}=-1, n_{2}=0$ & $3.5-3.5i$ & $\bm{42.0+8.0i}$ \\
    \hline
    $n_{1}=0, n_{2}=1$ & $-3.2+0.0i$ & $\bm{-29.0+0.0i}$\\
    \hline
    $n_{1}=0, n_{2}=-1$ & $-3.2-0.0i$  &  $0.08+3.21i$ \\
    \hline
    $n_{1}=1, n_{2}=1$ & $-3.2-0.0i$ & $0.21+3.2i$ \\ 
    \hline
    $n_{1}=-1, n_{2}=-1$ & $-3.2+0.0i$ & $\bm{-29.0+0.0i}$ \\
    \hline
\end{tabular}
\caption{Zeroth and first shell Fourier components of the intralayer potential ($l=l'$) and phase-modified top-to-bottom interlayer tunneling ($l\neq l'$) for $\boldsymbol{M}_{1}=\boldsymbol{b}_{1}/2$, rounded to 0.1 meV.
Together with the effective masses $m_{\perp,l}^{\boldsymbol{M}_{1}}=0.27 m_e$ and $m_{\parallel,l}^{\boldsymbol{M}_{1}}=2.41 m_e$, with $m_e$ the bare electron mass,
these values define the moir\'{e} model for $t$HfS$_2$ at valley $\boldsymbol{M}_{1}$.
\label{table:FourierComponents}}
\end{table}

\textit{Simplest models---} In 1T-HfS$_2$ crystalline bilayers, the intralayer potential terms are relatively small, and interlayer tunneling is dominated by the four boldface Fourier components in Table \ref{table:FourierComponents}. 
If we assume that the $\bd$-dependence in 
1T-MX$_{2}$ (M$=$Zr, Hf, Sn and X$=$S, Se) homobilayers is similar,
then for each valley, the lowest energy moir\'{e} conduction bands
are minimally described by a spin-independent two-orbital continuum model containing 
an intralayer kinetic term with anisotropic effective mass and 
stacking-dependent tunneling
$\Delta_{T}(\boldsymbol{M}_1;\bd)\approx
2\Re\left[e^{-i\boldsymbol{M}_{1}\cdot\bd}t_{C}\right]
+2\cos((\boldsymbol{M}_1+\boldsymbol{b}_{2})\cdot\bd)t_{R}$.
Here, $t_{C}\in\mathbb{C}$ and $t_{R}\in\mathbb{R}$ are material parameters
that can be deduced directly from DFT data
at three values of $\bd$, which we list in Supplemental Material \footnotemark[2].
The values of $t_{C}$ and $t_{R}$ account qualitatively for the 
values of $\bar{\Delta}_{\mathfrak{t}}^{\mathfrak{b}}(\boldsymbol{M}_{1};\bG)$
obtained by integrating over continuously varying stacking.

\textit{Moir\'e model---} 
We now address the $\bd$-independent 
dispersive term (see Eq.~(10) of the Appendix) in the moir\'e band model, which
is obtained by averaging the intralayer Hamiltonian matrix elements over $\bd$.
When described at parabolic order, the effective mass
tensor near valley $\boldsymbol{M}_{\nu}$ has eigenvectors in the $\bk$-space directions parallel ($a=\,\parallel$) and perpendicular ($a=\,\perp$) to $\boldsymbol{M}_{\nu}$.
The effective mass eigenvalues are weakly dependent on $\bd$ and are approximated
here by their spatial averages 
\footnote{Position-dependent effective masses can be included in the moir\'e band Hamiltonian when needed.}, 
denoted $m_{a,l}^{\boldsymbol{M}_{\nu}}$.
The explicit form of the plane-wave representation two-orbital continuum model 
Hamiltonian is
\begin{align}
    & \bra{l,\bq+\tilde{\boldsymbol{G}}} H_{\text{LD}}^{\boldsymbol{M}_{\nu}} \ket{l',\bq'+\tilde{\boldsymbol{G}'}} \nonumber\\
    &=\delta_{l,l'}\delta\big(\bq'-\bq\big)\Big(\delta_{\tilde{\bG}',\tilde{\bG}}\sum_{a\in\{\parallel,\perp\}}\frac{\hbar^2}{2m_{a,l}^{\boldsymbol{M}_{\nu}}}\big(\mathfrak{q}^{a}+\tilde{G}^{a}\big)^2 \nonumber\\
    &\qquad\qquad\qquad\qquad+\Delta_{l}(\boldsymbol{M}_{\nu};\tilde{\bG}'-\tilde{\bG})\Big) \nonumber\\
    &+(1-\delta_{l,l'})\delta\big(\bq'-\bq+(-1)^{\delta_{l,\mathfrak{t}}}\tilde{\bk}_{*}\big)
    \bar{\Delta}_{l'}^{l}(\boldsymbol{M}_{\nu};\tilde{\bG}'-\tilde{\bG})
    \label{Happrox1}
\end{align}
where $k^{\parallel}\equiv\bk\cdot\boldsymbol{M}_{\nu}/|\boldsymbol{M}_{\nu}|$ and $k^{\perp}\equiv\bk\cdot(\hat{\boldsymbol{e}}_{z}\cross\boldsymbol{M}_{\nu})/|\boldsymbol{M}_{\nu}|$.

\begin{figure}
	\centering
	\includegraphics[width= 1.0\linewidth]{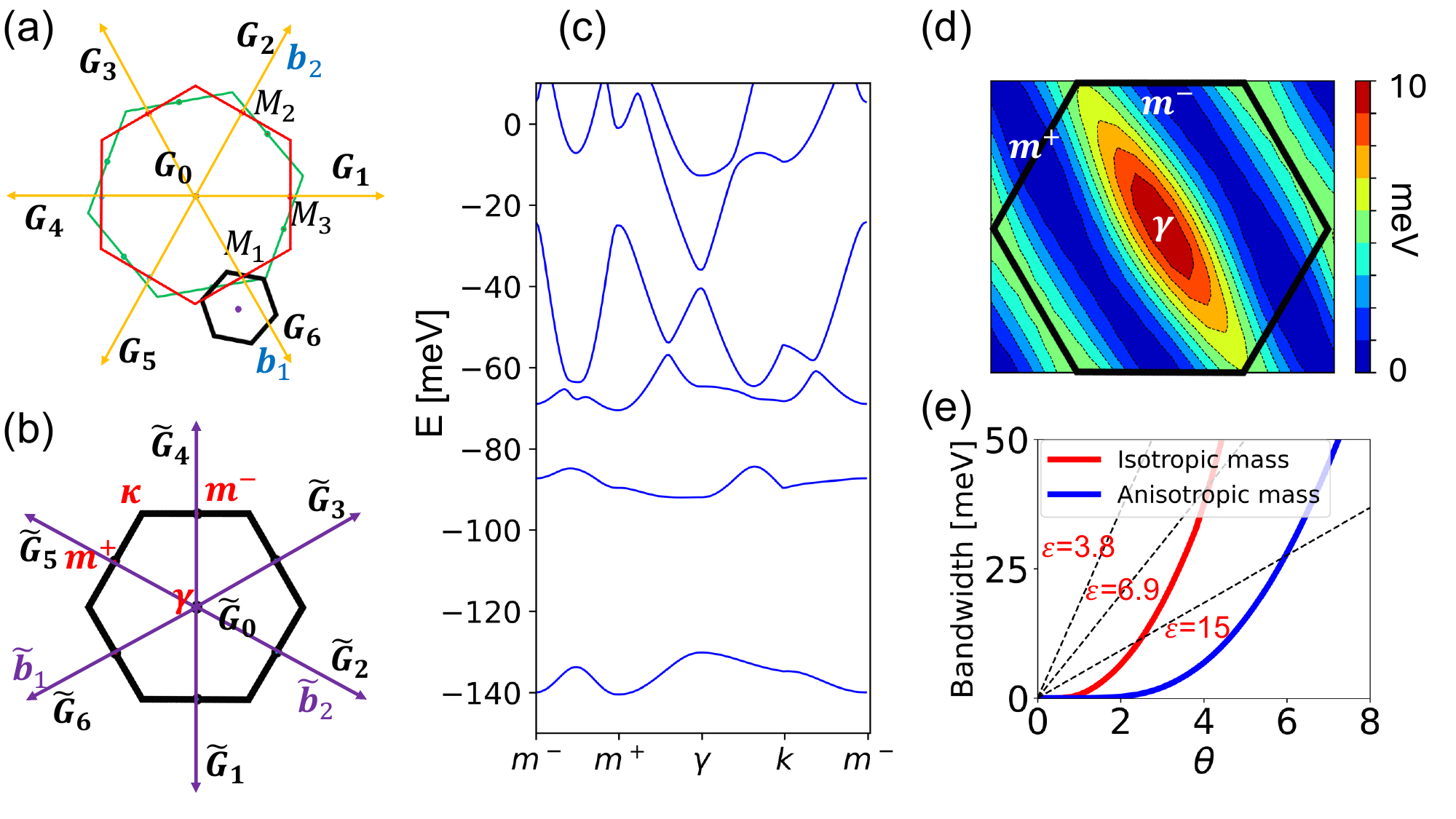}
	\caption{(a) Schematic illustration of the moir\'{e} BZ (black hexagon) in $t$HfS$_{2}$, where the red (green) hexagon illustrates the bottom (top) monolayer BZ. (b) Moir\'{e} ($\tilde{\boldsymbol{G}}_{i}$) and crystalline bilayer ($\boldsymbol{G}_{i}$) reciprocal lattice vectors. Points in mBZ defined relative to $\boldsymbol{M}_{1}$; $m^- = \boldsymbol{0}$, $m^+ = \tilde{\boldsymbol{M}}_{1}$,
    $\gamma =\tilde{\boldsymbol{G}}_1/2$, $\kappa = \gamma - (\tilde{\boldsymbol{G}}_1 + \tilde{\boldsymbol{G}}_2)/3$.
    (c) $\boldsymbol{M}_{1}$-valley moir\'e conduction bands for twist angle $\theta = 5^{\circ}$.
    (d) Contour plot of the lowest energy moir\'{e} band for $\theta = 5^{\circ}$
    and (e) its bandwidth for various values of $\theta$ (blue) compared to those in a hypothetical twisted bilayer with isotropic mass equal to the light  $m^{\boldsymbol{M}_{1}}_{\perp,l}$.
    The interaction energy scale $e^2/\epsilon a_M$ calculated with $\epsilon = 3.8,6.9,15$,
    for the out-of-plane and in-plane dielectric constants of bulk hBN and
    the out-of-plane HfS$_2$ dielectric constant, respectively,
    are plotted as dashed black curves.}
	\label{fig:moire_bands}
\end{figure}

\begin{figure}[t!]
	\centering
	\includegraphics[width= 0.7\linewidth]{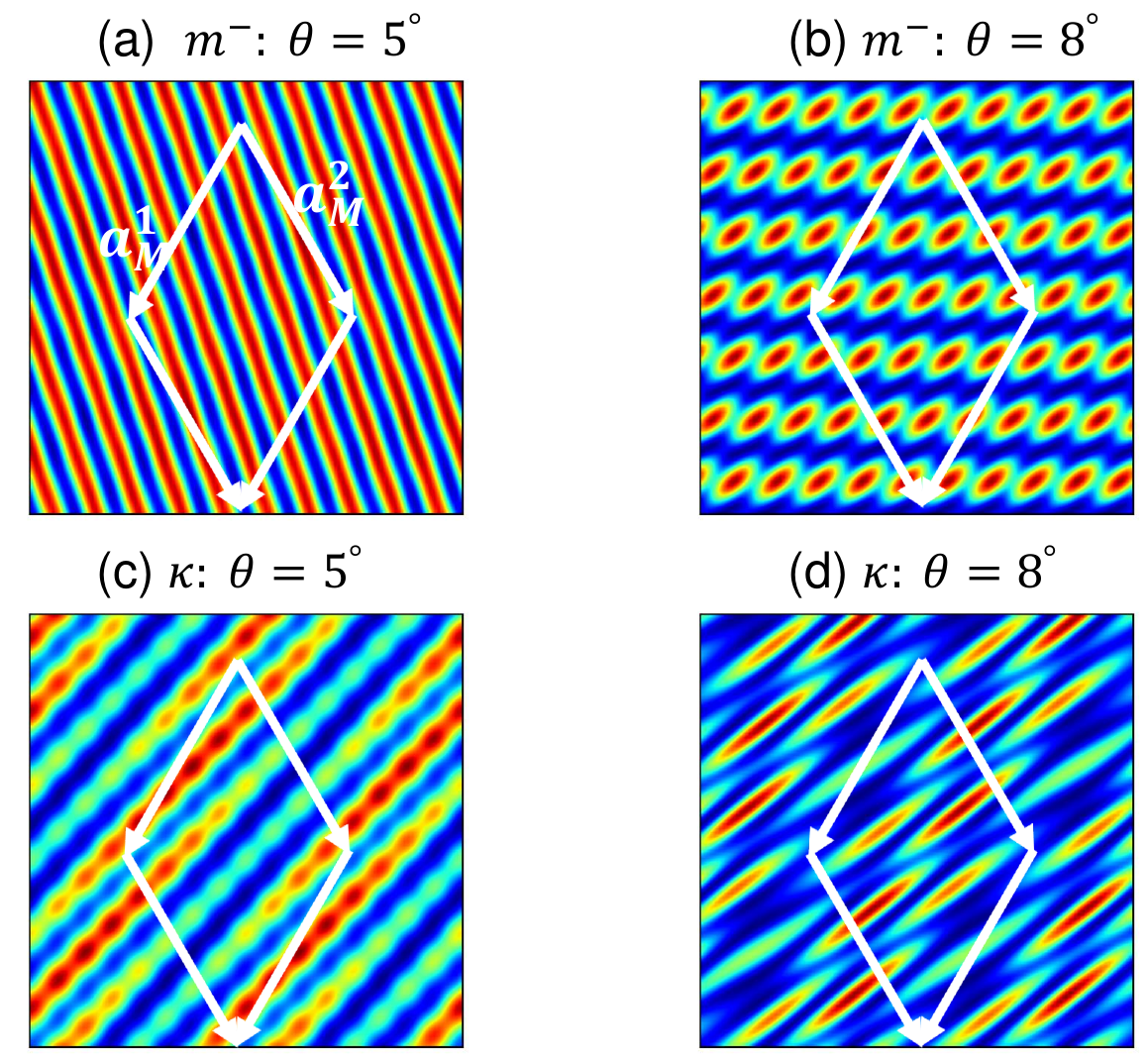}
	\caption{Spatial distribution of electron density
    in the lowest energy conduction band's moir\'{e} Bloch eigenfunction in $t$HfS$_{2}$ 
    for various moir\'e BZ momenta and twist angles. Parallelograms identify a moir\'{e} unit cell in each case.}
	\label{fig:contour}
\end{figure}

Our findings for the $\boldsymbol{M}_{1}$ valley are summarized in Fig.~\ref{fig:moire_bands},
and additional plots are included in Supplemental Material \footnotemark[2].
The effective masses in 1T-HfS$_2$ crystalline bilayers are highly anisotropic, as evidenced in Figs.~\ref{fig:Bands}(c)--(e): $m_{\parallel,l}^{\boldsymbol{M}_{\nu}}/m_{\perp,l}^{\boldsymbol{M}_{\nu}}\approx 9$.
As shown in Figs.~\ref{fig:moire_bands}(c) and \ref{fig:moire_bands}(d),
in the $\boldsymbol{M}_{1}$ valley, this leads to a lowest energy moir\'{e} conduction
band that is significantly more dispersive along the $\perp \boldsymbol{M}_{1}$ direction (which 
is along the $\tilde{\boldsymbol{b}}_{1}$ direction in the moir\'{e} BZ; 
see the $m^{-}$ to $m^{+}$ line)
and less dispersive along the $\parallel \boldsymbol{M}_{1}$ direction (see the $\kappa$ to $\gamma$ line).  
Up to an overall scaling factor related to variation in bandwidth, the shape of that band is qualitatively independent of $\theta$ for small twist angles \footnotemark[2].
However, in-plane strain relaxation in $t$HfS$_2$, whose influence
on key model parameters is discussed in the Appendix, can be important for $\theta\lesssim 4^{\circ}$ and lead to variation in the realized band structure \footnotemark[2].
Furthermore, Fig.~\ref{fig:contour} shows that the spatial distribution
of flat-band charge is more momentum-dependent within a band than in the case of
atomic crystals, which have strong attractive Coulombic potentials centered on nuclear positions.  This property is shared with other moir\'e material systems and
implies that interactions can affect not only the widths of bands but also their shapes as band filling factors change.  

\textit{Discussion---}
In this Letter, we employ a WF-based local displacement scheme \cite{MacDonald2014} to construct a moir\'{e} band model of the low-energy electronic conduction states in $t$HfS$_2$ and other group IV TMD homobilayers that have similar electronic structures.  
Our approach provides a universal moir\'e band Hamiltonian that applies at all small twist angles by evaluating displacement-dependent corrections to the Hamiltonian that are small on the atomic scale 
but significant on the moir\'e scale.  The low-energy single-particle Hilbert space separates into
six decoupled (flavor) subspaces labeled by spin ($s_z$) and valley ($\boldsymbol{M}_{\nu}$), each of which contains a layer degree of freedom.
We find (a) the moir\'e band Hamiltonian has strong valley-dependent effective mass anisotropy inherited from the host crystals, and a rotational symmetry implies that the large and small mass directions in different valleys are rotated in momentum space;
(b) a real-valued interlayer tunneling that splits layer-symmetric and layer-antisymmetric states and is 
peaked near metal on chalcogen layer configurations with a typical size $\sim 100$ meV;
and (c) potentials that are identical in the two layers and vary by $\sim 30$ meV over the 
moir\'e unit cell.  

Because the lowest energy moir\'e conduction bands are narrow and 
isolated, we expect that electronic properties in $M$-valley TMD moiré materials
will be controlled by strong correlation physics as they are in the 
$K$-valley systems studied previously.  
However, unlike in the $K$-valley case,
the moir\'{e} bands in $M$-valley TMDs will always have vanishing Chern numbers.
Other differences are readily anticipated.
Because the effective masses are larger in $M$-valley TMD systems,
we expect to see classical lattice-gas physics in which hopping between sites 
plays little role at small twist angles.  The large masses also 
imply band widths that remain smaller than typical interaction energy scales 
(see Fig.~\ref{fig:moire_bands}(e)) out to large twist angles, implying that   
strong correlations will persist to the twist angle range at which 
the complicating strain-relaxation effects \cite{zhang2024polarization} do not play a role.  This could raise ordering temperatures toward room temperature and simplify comparisons between theory and experiment, in spite of the high native dielectric constants of the materials of interest.  Spin-valley flavor magnetism appears likely to be at least as common 
in $M$-valley TMDs as it is in $K$-valley TMDs, since the underlying exchange interactions 
are weakened by nontrivial band topology.  
This can lead to Fermi surface reconstructions and interaction-induced insulating states at all 
integer band fillings $\nu$ smaller than the filling $\nu=6$ at which the lowest band is fully filled.
Moreover, it appears that there
are close analogies between multivalley spin-order physics in the insulating states of moir\'e materials and the multiorbital ordering common in many 
insulating transition metal oxides with active $t_{2g}$ orbitals 
\footnote{Within the subspace of $t_{2g}$ states in transition metal oxide perovskites, the dominant hopping processes conserve orbital type just as 
the low-energy conduction states in the M-valley moir\'{e}s considered here can be described by a moir\'{e}-scale tight-binding model with hoppings that conserve valley.
Hopping between particular lattice sites is strongly orbital-dependent in the perovskite case and valley-dependent in the moir\'e case.  See, e.g., Sec.~5 of Ref.~\cite{kugel1982jahn}.}.
The mismatch in shape between Fermi surface contours in different valleys
impacts the competition \cite{li2016} between valley polarized and intervalley coherent states.  These two ordered states are readily distinguished
in transport experiments because of the valley-dependent mass anisotropy. 
It also seems likely that the strongly anisotropic bands may bring the tendency toward density-wave states and other aspects of quasi-one-dimensional material physics to the moir\'e platform.
Although we are confident in many of these predictions, we do expect that experimental 
studies of strongly correlated $M$-valley moiré materials will have their own surprises.
One particularly intriguing question is whether superconductivity will appear as 
regularly near magnetic phase transitions as it does in the $K$-valley case.

\textit{Note Added---}
As this Letter was being prepared for publication, we learned of a closely related study by Călugăru {\it et al.}~\cite{princetonref}, which reaches similar conclusions.

\textit{Acknowledgments---}
We are grateful to Andrei Bernevig, Dumitru Călugăru, Yi Jiang, Haoyu Hu, Hanqi Pi and collaborators for 
insightful discussions near the conclusion of this work.
We acknowledge HPC resources provided by
the Texas Advanced
Computing Center at The University of Texas at Austin.
This work was supported by a Simons Foundation Collaborative Research Grant
and by the Robert A.~Welch Foundation under Grant Welch F-2112.

\bibliography{moireHfS2}

\onecolumngrid
\begin{center}
{\bf{End Matter}}
\end{center}
\twocolumngrid
\textit{Appendix: Moir\'{e} band models via the local displacement scheme---}
The local displacement approach \cite{MacDonald2014} is an approximation scheme that can be used to construct accurate periodic effective models of the 
low-energy (electronic) states in semiconductor bilayers that have small $\theta$ and/or $\lambda$ close to $1$,
and weak coupling between layers.
This approach employs a family of Bloch Hamiltonians 
$H(\boldsymbol{\mathsf{d}})$ 
for (fictitious) crystalline insulator bilayers --
constructed as aligned stacks of the quasicrystal's constituent layers, but with lattice constants (of the top layer, for instance) that are artificially adjusted (if necessary) so that the top and bottom layers are characterized by the same Bravais lattice $\Lambda$
\footnote{For each $\bd\in\mathbb{R}^{2}$, the Bloch Hamiltonian $H(\boldsymbol{x},\boldsymbol{p}(\boldsymbol{x});\boldsymbol{\mathsf{d}})$ of the crystal bilayer is $\Lambda$-periodic in $\boldsymbol{x}$, where $\Lambda$ is the Bravais lattice of the (arbitrarily chosen) bottom layer.
We choose $\Lambda$ and its real-space $\Omega_{uc}$ (reciprocal space BZ) unit cell to be independent of $\bd$.
The vertical separation between layers is fixed at each $\bd$ by minimizing energy.} --
that is parametrized by an in-plane relative displacement $\bd\in\mathbb{R}^2$ between layers.
Our approach relies on identifying a set of Bloch-type
vectors $\big|\tilde{\phi}_{I,\bk}(\bd)\big\rangle$ such that
for each $\bd$ and each $\bk$ near VBM and/or CBM (the set of which we denote $\mathcal{K}\subset\text{BZ}$), $\big\{\big|\tilde{\phi}_{I,\bk}(\bd)\big\rangle\big\}_{I}$
spans the low-energy Hilbert space, and that are smooth over BZ such that they transform into localized orbitals \cite{Brouder2007}.
We require, in addition, that those orbitals are localized in either the top or bottom layer.
One potential way to achieve this, which is natural in weakly coupled bilayers, is to construct WFs \cite{Marzari2012}.
AA-stacked crystalline homobilayers of 1T-MX$_2$ have TRS, and the set of six doubly-degenerate energy bands just above the band gap at charge neutrality, to which we restrict our focus,
remains isolated for all $\bd$ \footnotemark[2].
Thus, for each $\bd$, a set of WFs $\ket{W_{I,\boldsymbol{R}}(\boldsymbol{\mathsf{d}})}$
can be constructed \cite{Brouder2007} for those bands. 
These WFs can be chosen such that the label $I$ decomposes into layer- and orbital-type (which includes orbital structure and spin) labels $I=(l,\alpha_{l})$ \footnotemark[2], where 
$l=\mathfrak{b}$ and $l=\mathfrak{t}$ identify bottom and top layers, respectively;
indeed, the identification of 
layer as a degree of freedom is essential and is the main advantage of a Wannier-based approach.  
Thus, the Bloch-type vectors $\big|\tilde{\psi}_{(l,\alpha_l),\boldsymbol{k}}(\boldsymbol{\mathsf{d}})\big\rangle$ defined by the inverse Bloch-Floquet transform \cite{Marzari2012} of the WFs $\ket{W_{(l,\alpha_l),\boldsymbol{R}}(\boldsymbol{\mathsf{d}})}$ satisfy the aforementioned criteria.
The Hamiltonian matrix elements
\begin{align}
    \mathscr{H}^{l,\alpha_{l}}_{l',\beta_{l'}}(\boldsymbol{k};\boldsymbol{\mathsf{d}})
    \equiv\bra{\tilde{\psi}_{(l,\alpha_{l}),\boldsymbol{k}}(\boldsymbol{\mathsf{d}})}H(\boldsymbol{\mathsf{d}})\ket{\tilde{\psi}_{(l',\beta_{l'}),\boldsymbol{k}}(\boldsymbol{\mathsf{d}})}
    \label{PseudomatEl1}
\end{align}
contain microscopic data needed to construct the moir\'{e} band model.

It is commonly assumed, and it is the case here \footnotemark[2], that the WFs can be chosen to track smoothly with $\bd$ so that $W_{(l,\alpha),\boldsymbol{R}}(\boldsymbol{r};\boldsymbol{\mathsf{d}}+\boldsymbol{\mathsf{a}})\approx W_{(l,\alpha),\boldsymbol{R}}(\boldsymbol{r}-\delta_{l,\mathfrak{t}}\boldsymbol{\mathsf{a}};\boldsymbol{\mathsf{d}})$
after a relative translation of layers by $\boldsymbol{\mathsf{a}}\in\mathbb{R}^{2}$.
It follows that  
\begin{align}
    \ket{\tilde{\psi}_{(l,\alpha),\boldsymbol{k}}(\bd+\boldsymbol{\mathsf{a}})}\approx 
    e^{-i\delta_{l,\mathfrak{t}}\boldsymbol{k}\cdot\boldsymbol{\mathsf{a}}}
    \ket{\tilde{\psi}_{(l,\alpha),\boldsymbol{k}}(\boldsymbol{\mathsf{d}})},
\end{align} 
and therefore, that 
\begin{align}
	\bar{\mathscr{H}}^{l,\alpha_{l}}_{l',\beta_{l'}}(\boldsymbol{k};\boldsymbol{\mathsf{d}})\equiv
	e^{i(\delta_{l',\mathfrak{t}}-\delta_{l,\mathfrak{t}})\boldsymbol{k}\cdot\boldsymbol{\mathsf{d}}}\mathscr{H}^{l,\alpha_{l}}_{l',\beta_{l'}}(\boldsymbol{k};\boldsymbol{\mathsf{d}})
	\label{PseudomatEl}
\end{align}
is $\Lambda$-periodic in $\bd$ and admits a Fourier series expansion with coefficients given by
\begin{align}
	\bar{\mathscr{H}}^{l,\alpha_{l}}_{l',\beta_{l'}}(\boldsymbol{k};\boldsymbol{G})=
	\frac{1}{\Omega_{uc}}\int_{\Omega_{uc}}e^{i\boldsymbol{G}\cdot\boldsymbol{\mathsf{d}}}\bar{\mathscr{H}}^{l,\alpha_{l}}_{l',\beta_{l'}}(\boldsymbol{k};\boldsymbol{\mathsf{d}}) \; d\boldsymbol{\mathsf{d}}
    \label{FourierTransf}
\end{align}
for $\boldsymbol{G}\in\Lambda^{*}$.
Here, bar accents are used to identify phase-modified quantities.

There are two steps in this scheme \cite{MacDonald2014}.
First, in each matrix element
$\bra{W_{(l,\alpha_{l}),\boldsymbol{R}}(\boldsymbol{\mathsf{d}})}H(\boldsymbol{\mathsf{d}})\ket{W_{(l',\beta_{l'}),\boldsymbol{R}'}(\boldsymbol{\mathsf{d}})}$,
$\bd$ is replaced by $\boldsymbol{d}(\boldsymbol{r})$, the local stacking vector at position $\boldsymbol{r}$ in the moir\'e pattern.
For the actual quasicrystalline structure, one can define \cite{cances2017generalized,Luskin2017} 
a function of lattice sites
$\boldsymbol{d}_{l}:\Lambda_{l}\rightarrow \Omega_{uc}^{\bar{l}}$ to encode
the distance between layer $l$ lattice site $\boldsymbol{R}$ and the closest lattice site in the opposite layer ($\bar{l}$)
\footnote{The top (bottom) layer of the quasicrystal has Bravais lattice $\Lambda_{\mathfrak{t}}$ ($\Lambda_{\mathfrak{b}}$) and
$\Lambda_{\mathfrak{t}}=\lambda R_{\theta}^{z}(\Lambda_{\mathfrak{b}})$.
For small $\theta$ and $\lambda-1$,
and for $\boldsymbol{R}\in\Lambda_{\mathfrak{b}}$,
$\lambda R_{\theta}^{z}(\boldsymbol{R})-\boldsymbol{R}\approx(\lambda-1)\boldsymbol{R}+\theta\hat{\boldsymbol{e}}_{z}\cross\boldsymbol{R}$. Modulating this vector by $\Lambda_{\mathfrak{t}}$ produces a position in the top layer unit cell and encodes the relative position of the nearest element of $\Lambda_{\mathfrak{t}}$ to $\boldsymbol{R}\in\Lambda_{\mathfrak{b}}$.
Analogous arguments follow for the top layer.
Since we choose $\Lambda\equiv\Lambda_{\mathfrak{b}}$, even when $l=\mathfrak{t}$ in the Hamiltonian matrix elements $\boldsymbol{R}\in\Lambda$ is not generically in $\Lambda_{\mathfrak{t}}$. An additional lattice-site-dependent position shift of top layer WFs in the displaced bilayer must therefore be included when approximating quasicrystal matrix elements that involve layer $\mathfrak{t}$.
}. 
$\boldsymbol{d}(\boldsymbol{r})$ is the coarse-grained version of $\boldsymbol{d}_{l}(\boldsymbol{R})$
which, for small $\theta$ and $\lambda-1$, varies slowly on the lattice scale and is  
chosen to vary continuously rather than being folded discontinuously 
to a lattice primitive cell.
It follows that 
$\boldsymbol{d}(\boldsymbol{r})\approx(\lambda-1) \, \boldsymbol{r}+\theta \, \hat{\boldsymbol{e}}_{z}\cross\boldsymbol{r}$
with $\hat{\boldsymbol{e}}_{z}$ the surface-normal direction.
So far, we have neglected in-plane strain relaxation relative to rigid rotation, which can be important, but this correction can be added separately when needed \cite{jung2015origin}. The effect of in-plane relaxation is primarily to make the local potential and tunneling parameters vary more quickly in the domain walls that form when the moir\'e periods are long and less quickly outside the domain walls.  Since these spatial variations of these processes enter the continuum model via their Fourier components, higher Fourier components are needed when the domain wall widths are 
much smaller than the moir\'e periods.  For TMDs, these continuum model
elaborations are quantitatively significant for twist angles below a few degrees.  Secondly, the value of the 
$l=l'$, $\boldsymbol{G}\neq\boldsymbol{0}$ and $l\neq l'$
terms at $\bk$ is approximated by that at the nearby valley $\bk_{*}\in\mathcal{K}$.
Both elements are justified \cite{MacDonald2014} for 
sufficiently long moir\'e periods because the $\bk$-dependence of the 
Hamiltonian at each $\bd$ is on scale $\Lambda^*$.

With these approximations, the local displacement 
Hamiltonian $H_{\text{LD}}^{\boldsymbol{k}_{*}}$ for 
valley $\boldsymbol{k}_{*}\in\mathcal{K}$ acts 
in an envelope function space \cite{CardonaBook} 
that is the direct product of layer, Wannier orbital (within each layer), and 
continuous position subspaces, and its eigenstates are spinors with 
layer and Wannier orbital indices.  Using a momentum-space representation 
for the position degree of freedom,
    \begin{align}
    &\bra{l,\alpha_{l},\boldsymbol{k}} H_{\text{LD}}^{\boldsymbol{k}_{*}} \ket{l',\beta_{l'},\boldsymbol{k}'} \nonumber\\
    &= \delta_{l,l'}\Big(\delta\big(\boldsymbol{k}'-\boldsymbol{k}\big)\mathscr{H}^{l,\alpha_{l}}_{l,\beta_{l}}(\boldsymbol{k};\boldsymbol{G}=\boldsymbol{0})\nonumber\\
    &\qquad\qquad+\sum_{\boldsymbol{G}\in\Lambda^{*}\setminus\{\boldsymbol{0}\}}
    \delta\big(\boldsymbol{k}'-\boldsymbol{k}-\tilde{\boldsymbol{G}}\big)\mathscr{H}^{l,\alpha_{l}}_{l,\beta_{l}}(\boldsymbol{k}_{*};\boldsymbol{G})\Big)\nonumber\\
    &+(1-\delta_{l,l'})\sum_{\boldsymbol{G}\in\Lambda^{*}}\delta\big(\boldsymbol{k}'-\boldsymbol{k}+(-1)^{\delta_{l,\mathfrak{t}}}\tilde{\bk}_{*}-\tilde{\boldsymbol{G}}\big)\bar{\mathscr{H}}^{l,\alpha_{l}}_{l',\beta_{l'}}(\boldsymbol{k}_{*};\boldsymbol{G}).
    \label{Happrox}
    \end{align}
Here,
$\tilde{\boldsymbol{G}}\equiv\lambda R^{z}_{-\theta}(\boldsymbol{G})-\boldsymbol{G}$ 
are elements of the moir\'{e} reciprocal lattice $\Lambda_{\text{M}}^{*}$ 
\footnote{$\Lambda_{\text{M}}^{*}\equiv\{\lambda R_{-\theta}(\boldsymbol{G})-\boldsymbol{G}:\boldsymbol{G}\in\Lambda^{*}\}\subset\mathbb{R}^{2}$ is a Bravais lattice called the moir\'{e} reciprocal lattice. Its dual, the moir\'{e} lattice, is given by $\Lambda_{\text{M}}=\{\boldsymbol{r}\in\mathbb{R}^{2}:\boldsymbol{d}(\boldsymbol{r})\in\Lambda\}$.},
$\tilde{\boldsymbol{k}}_{*}\equiv\lambda R^{z}_{-\theta}(\boldsymbol{k}_{*})-\boldsymbol{k}_{*}$,
and $\bk$ and $\bk'$ are extended from BZ to $\mathbb{R}^{2}$.
Because the layer separations in the van der Waals crystals from which moir\'e materials can be successfully fabricated are always much larger than the lattice constants within layers,
sizable values for $\bar{\mathscr{H}}^{l,\alpha_{l}}_{l',\beta_{l'}}(\boldsymbol{k};\boldsymbol{G})$ are expected for $\boldsymbol{G}$
only in the first few shells of $\Lambda^{*}$, implying that the
lowest energy moir\'{e} bands involve decoupled pockets of momentum space that are small compared to the crystalline Brillouin zone size.  This property justifies the valley decoupling that occurs in the 
local displacement approximation and also justifies an expansion of $\bar{\mathscr{H}}^{l,\alpha_{l}}_{l,\beta_{l}}(\boldsymbol{k};\boldsymbol{G}=\boldsymbol{0})$ around the CBM 
$\boldsymbol{k}_{*}$ that is parametrized by an effective mass tensor. 
A key property of the $M$-valley TMDs considered here is that their effective mass tensors are anisotropic. 

We have neglected in-plane strain relaxation, but this can be included
by modifying the $\boldsymbol{d}:\mathbb{R}^{2}\rightarrow\mathbb{R}^{2}$
mapping between lattice sites and position from the linear rigid-displacement form
to the form that minimizes the sum of elastic and interlayer interaction energies \cite{jung2015origin}.
The effect of in-plane relaxation (we include vertical relaxation) is 
primarily to make the local potential and tunneling parameters vary more 
quickly in the domain walls that form when the moir\'e periods are long and 
less quickly outside the domain walls.  Since these spatial variations of these processes
enter the continuum model via their Fourier components, higher 
Fourier components are needed when the domain wall widths are 
much smaller than the moir\'e periods.  For TMDs, these continuum model
elaborations are quantitatively significant for twist angles below a few degrees.  
In the family of materials studied here, the relatively large 
effective masses mean that the most interesting many-body physics will occur at larger
twist angles where lateral relaxation is unimportant.
Although strain effects can be important quantitatively, 
the qualitatively new aspects of $M$-valley TMD moir\'{e} materials identified here,
due to TRIM valleys and effective mass anisotropy, 
will persist in the presence of strain.

\end{document}


\title{Supplementary Material for \\``Moir\'{e} Band Theory of M-Valley Twisted Transition Metal Dichalcogenides''}

\author{Chao Lei}
\email{leichao.ph@gmail.com}

\author{Perry T.~Mahon}
\email{perry.mahon@austin.utexas.edu}

\author{A.~H.~MacDonald}
\affiliation{Department of Physics, University of Texas at Austin, Austin, Texas 78712, USA}

\thanks{C.~L.~and P.~T.~M.~contributed equally to this work.}

\maketitle

\newpage

\section{Ab initio results for rigidly displaced AA-stacked crystalline homobilayers}

Density functional theory calculations were performed using Quantum Espresso \cite{QE2009} with pseudopotentials generated using the Perdew-Burke-Ernzerhof (PBE) exchange-correlation functional within the Projector Augmented Wave (PAW) method.
For each stacking the relaxation of bilayers is performed along the surface-normal ($\hat{\mathbf{z}}$) direction, with the positions of transition metal atoms fixed in the $x$--$y$ plane. During the relaxation, convergence threshold on total energy for ionic minimization is set to be $10^{-8}$ in atomic units of energy, while convergence threshold on forces for ionic minimization is set to be $10^{-8}$ in atomic units. The kinetic energy cutoff for wavefunctions is set to be 80 Ry and kinetic energy cutoff for charge density and potential is set to be 320 Ry. Methfessel-Paxton first-order spreading is used in the smearing with value of the gaussian spreading for Brillouin-zone integration as 0.02 Ry. Semi-empirical Grimme's DFT-D2 is used for the van der Waals correction. A 20 \AA thick vacuum was inserted to avoid the interactions between the images. The structure is relaxed with BFGS quasi-Newton algorithm. k-point sampling is set to be $8 \times 8 \times 1 $ during the relaxation and $12 \times 12 \times 1 $ during the self-consistent calculations. The lattice constants for 1T-TMDs used in the calculations are shown in Table \ref{tab:lat_const}.

In this section we plot various quantities as a function of relative in-plane layer displacement $\bd$. We adopt the convention that Bravais lattice basis vector $\boldsymbol{a}_{1}$ ($\boldsymbol{a}_{2}$) of the crystalline bilayer -- which is chosen independent of $\bd$ (see main text) -- is parallel to the bottom (left) boundary line of each such plot. The plot range in each basis vector direction is $0$ to $1$ in units of $|\boldsymbol{a}_{1}|=|\boldsymbol{a}_{2}|$.

\begin{table}[h!]
    \centering
    \begin{tabular}{||c|cccccc||}
    \hline
        Name & HfS$_2$ & HfSe$_2$ & ZrS$_2$ & ZrSe$_2$ & SnS$_2$ & SnSe$_2$ \\
        \hline
        a (\AA) & 3.64 & 3.74 & 3.66 & 3.77 & 3.70 & 3.86 \\
        \hline
        h (\AA) & 3.21 & 3.23 & 2.90 & 3.16 & 2.96 & 3.19 \\
        \hline
    \end{tabular}
    \caption{Lattice constants $a$ of 1T-TMDs and vertical distance $h$ between chalcogen atoms in top and bottom layers.}
    \label{tab:lat_const}
\end{table}

\subsection{$\bd$-dependence of AA-stacked crystalline bilayer 1T-HfS$_2$}
\begin{figure}[hbt!]
	\centering
	\includegraphics[width= 0.85\linewidth]{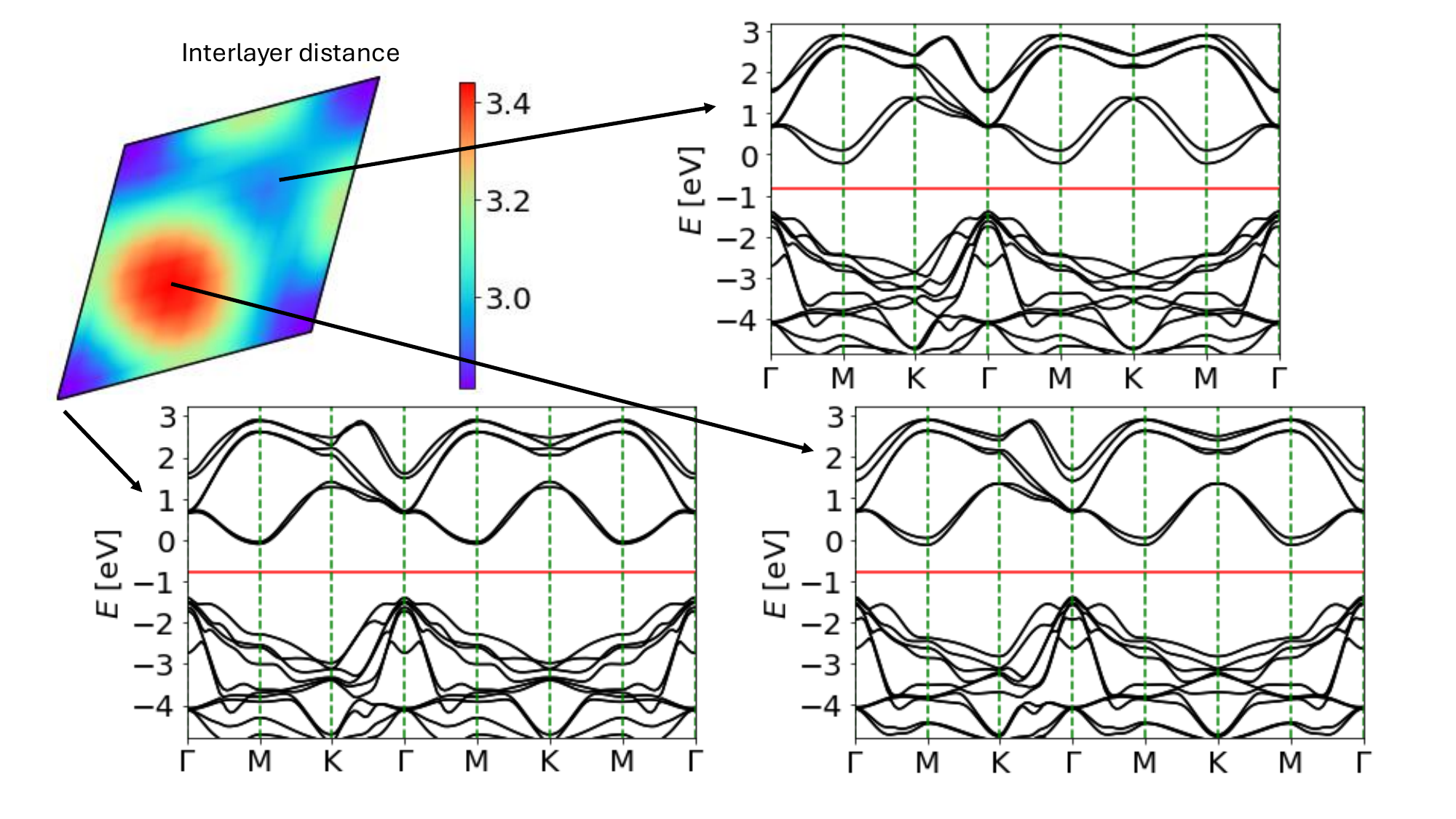}
	\caption{(top left) Interlayer distance (\r{A}) {\it vs.}~relative in-plane displacement $\bd$ of the constituent layers in aligned crystalline bilayer 1T-HfS$_2$. DFT bandstructure for $\bd=\boldsymbol{0}$ (bottom left), $\bd=\boldsymbol{a}_{1}/3+\boldsymbol{a}_{2}/3$ (bottom right), and $\bd=2\boldsymbol{a}_{1}/3+2\boldsymbol{a}_{2}/3$ (top right). 
    These bandstructures are representative of the findings for all $\bd$. 
    In particular, the six doubly-degenerate lowest energy conduction bands remain energetically isolated from all other bands at each $\bk\in\text{BZ}$ and vary weakly with $\bd$. In the main text we present a six-orbital Wannier tight-binding model (for each of the identical spin-$s_z$ sectors) of these conduction Bloch states at $\bd$, the bandstructure of which (see red bands in Fig.~2(c)--(e) of the main text) is faithful to these results.}
    \label{appendix:fig1}
\end{figure}

\begin{figure}[!htb]
	\centering
	\includegraphics[width= 0.35\linewidth]{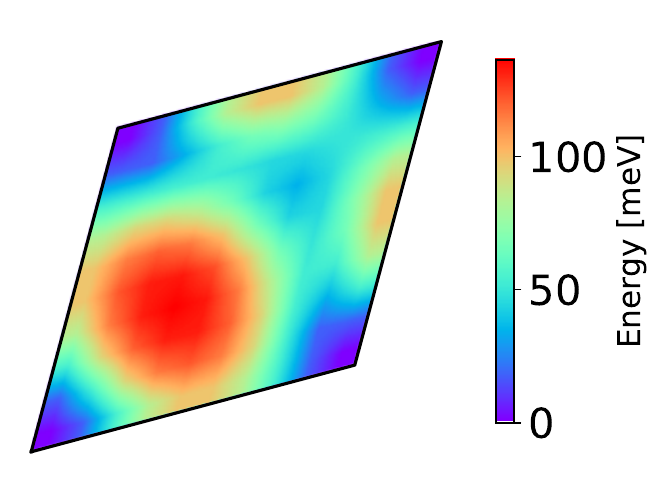}
	\caption{
    Total energy of crystalline bilayer 1T-HfS$_2$ as a function of in-plane displacement $\bd$ relative to placing the aligned 2D crystals directly on top of each other. 
    For $\bd=2(\boldsymbol{a}_{1}+\boldsymbol{a}_{2})/3$ the metal atom of the top layer is above the top chalcogen in the layer below. For the high energy $\bd=(\boldsymbol{a}_{1}+\boldsymbol{a}_{2})/3$ stacking neither metal atom is directly above or below a chalcogen of the adjacent layer.
    This energy scale is used to estimate the twist angle below which in-plane relaxation cannot be neglected. The difference in energy per metal atom between the low-energy thermodynamically stable $\bd=\boldsymbol{0}$ and the high energy $\bd=(\boldsymbol{a}_{1}+\boldsymbol{a}_{2})/3$ stackings is 
    $\Delta V=136$ meV.}
	\label{fig:SI_energyLandscape}
\end{figure}

\begin{figure}
    \centering
    \subfigure[]{\includegraphics[width=0.32\textwidth]{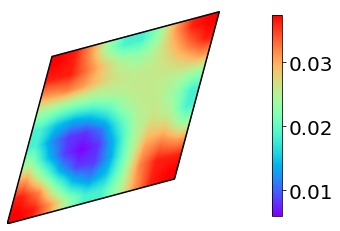}} 
    \subfigure[]{\includegraphics[width=0.32\textwidth]{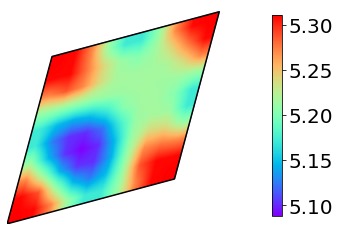}}
    \caption{As a function of relative in-plane layer displacement $\bd$ in the crystal unit cell, we plot for one of the three bottom layer WFs: (a) the distance (\r{A}) between that WF's center and the bottom layer Hf site position in the corresponding unit cell, and (b) its spread ($\text{\r{A}}^2$).
    Although this WF's center and its spread are not constant in $\bd$, the variation in both quantities is very small as $\bd$ is varied over the crystal unit cell.
    We therefore conclude that $W_{(l,\alpha),\boldsymbol{R}}(\boldsymbol{r};\boldsymbol{\mathsf{d}}+\boldsymbol{\mathsf{a}})\approx W_{(l,\alpha),\boldsymbol{R}}(\boldsymbol{r}-\delta_{l,\mathfrak{t}}\boldsymbol{\mathsf{a}};\boldsymbol{\mathsf{d}})$.}
\end{figure}

\begin{figure}[hbt!]
	\centering
	\includegraphics[width= 0.9\linewidth]{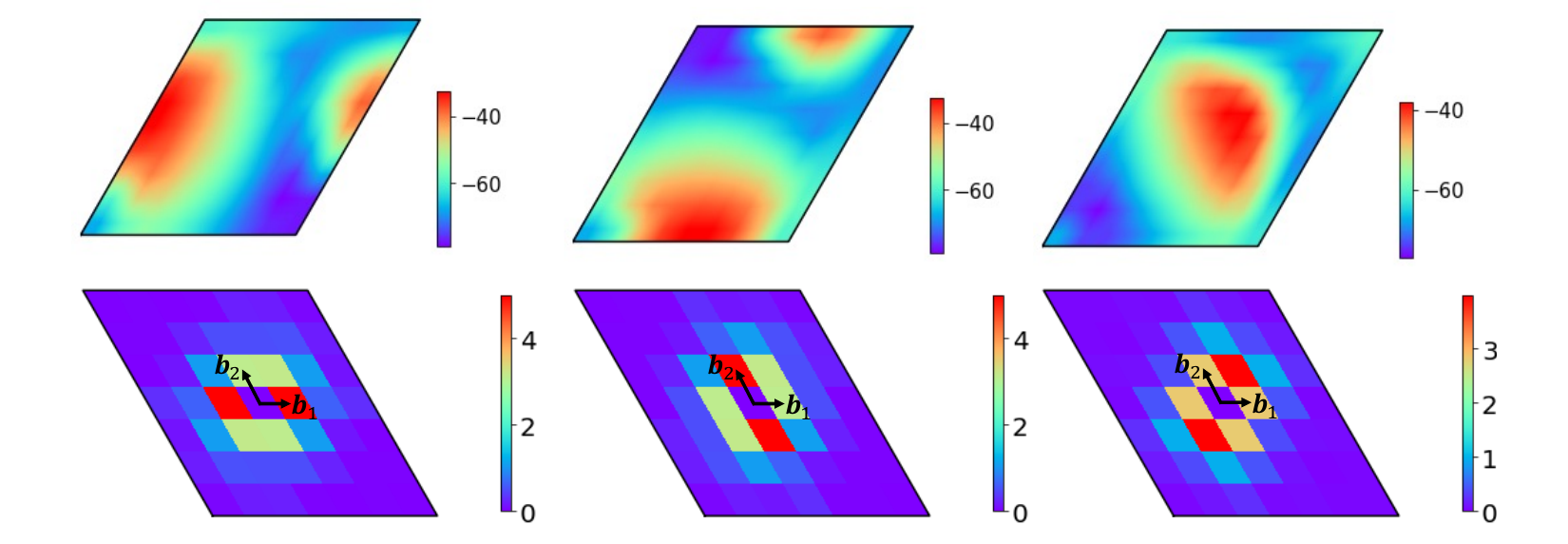}
	\caption{(top row) Intralayer potential $\Delta_{l}(\boldsymbol{M}_{\nu};\bd)$ {\it vs.}~stacking $\bd$ at the three inequivalent valleys (left to right: $\boldsymbol{M}_{1}$, $\boldsymbol{M}_{2}$, $\boldsymbol{M}_{3}$). As described in the main text, the valleys being at time-reversal momenta and the system having inversion symmetry together implies $\Delta_{\mathfrak{t}}(\boldsymbol{M}_{\nu};\bd)=\Delta_{\mathfrak{t}}(\boldsymbol{M}_{\nu};\bd)$. (bottom row) Magnitude of the Fourier components $\Delta_{l}(\boldsymbol{M}_{\nu};\boldsymbol{G})$, which decay quickly to zero with increasing $|\boldsymbol{G}|$.}
	\label{fig:SI_FourierPptential}
\end{figure}

\begin{figure}[!htb]
	\centering
	\includegraphics[width= 0.75\linewidth]{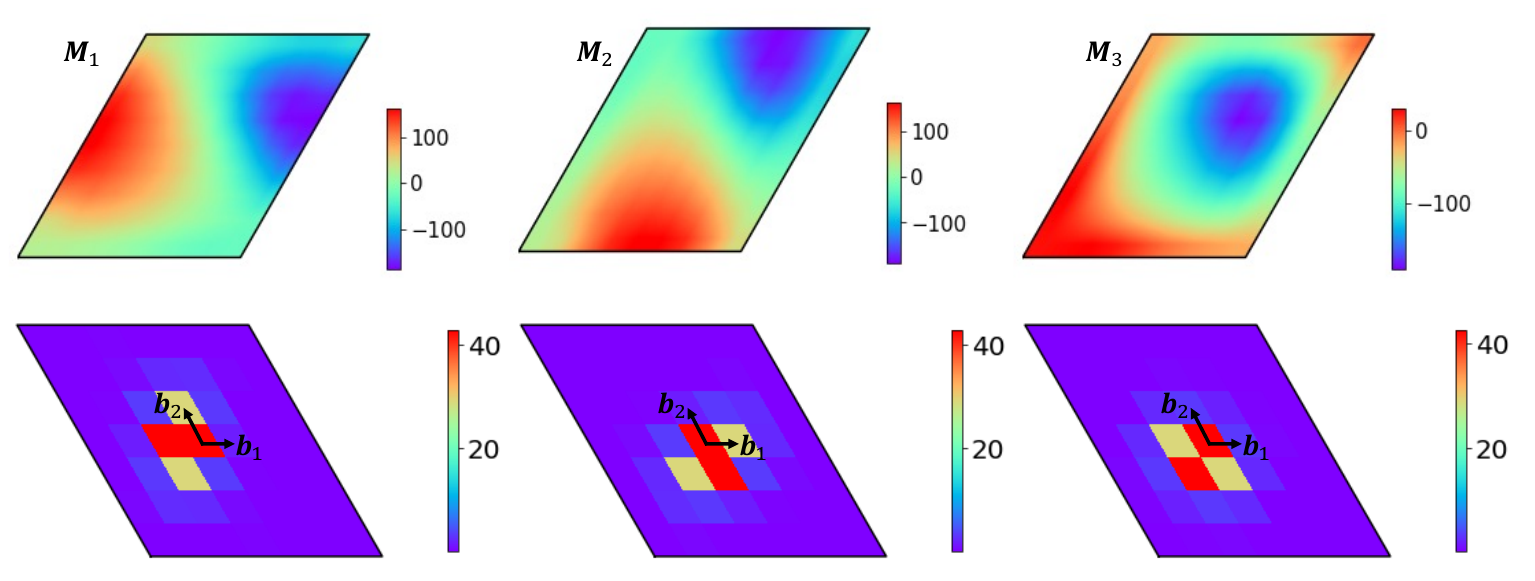}
	\caption{(top row) Interlayer tunneling $\Delta_{T}(\boldsymbol{M}_{\nu};\bd)$ {\it vs.}~stacking $\bd$ at the three inequivalent valleys. 
    The sign is chosen such that the lowest energy CBM state has layer parity consistent with that obtained from DFT (see Table \ref{table:bandsplitting}).
    For example, for $\bd=\boldsymbol{0}$ we find at each valley that the lowest energy CBM state has odd layer parity; therefore, we take $\Delta_{T}(\boldsymbol{M}_{\nu};\bd=\boldsymbol{0})>0$.
    As described in the main text, $\Delta_{T}(\boldsymbol{M}_{\nu};\bd)$ is quasi-$\Lambda$-periodic in $\bd$. 
    We define the $\Lambda$-periodic top-to-bottom interlayer tunneling $\bar{\Delta}_{\mathfrak{t}}^{\mathfrak{b}}(\boldsymbol{M}_{\nu};\bd)\equiv e^{i\boldsymbol{M}_{\nu}\cdot\bd}\Delta_{T}(\boldsymbol{M}_{\nu};\bd)$ for valley $\boldsymbol{M}_{\nu}$. 
    (bottom row) Magnitude of Fourier components $\bar{\Delta}_{\mathfrak{t}}^{\mathfrak{b}}(\boldsymbol{M}_{\nu};\boldsymbol{G})$, which decay quickly to zero with increasing $|\boldsymbol{G}|$.}
	\label{fig:SI_FourierTunnel}
\end{figure}

\FloatBarrier

\subsection{$\bd$-dependence of other proposed candidate group IV moir\'{e} TMDs}
Implementing similar DFT calculations performed above for crystalline bilayers of 1T-HfS$_{2}$, 
we study the $\bd$-dependence of AA-stacked crystalline homobilayers consisting other proposed candidate group-IV $M$-valley TMDs.

\begin{figure}[hb!]
    \centering
    \subfigure[]{\includegraphics[width=0.32\textwidth]{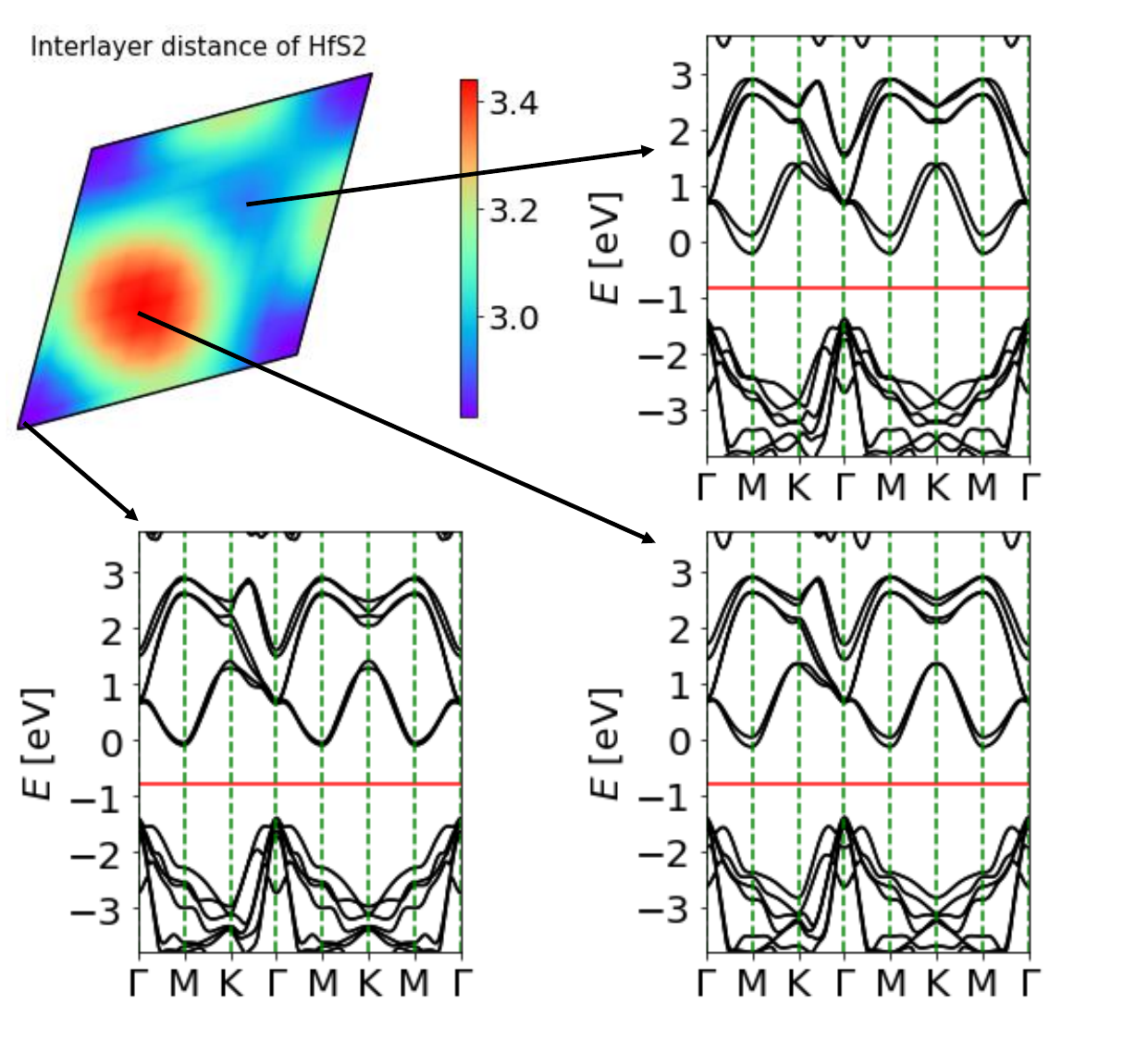}} 
    \subfigure[]{\includegraphics[width=0.32\textwidth]{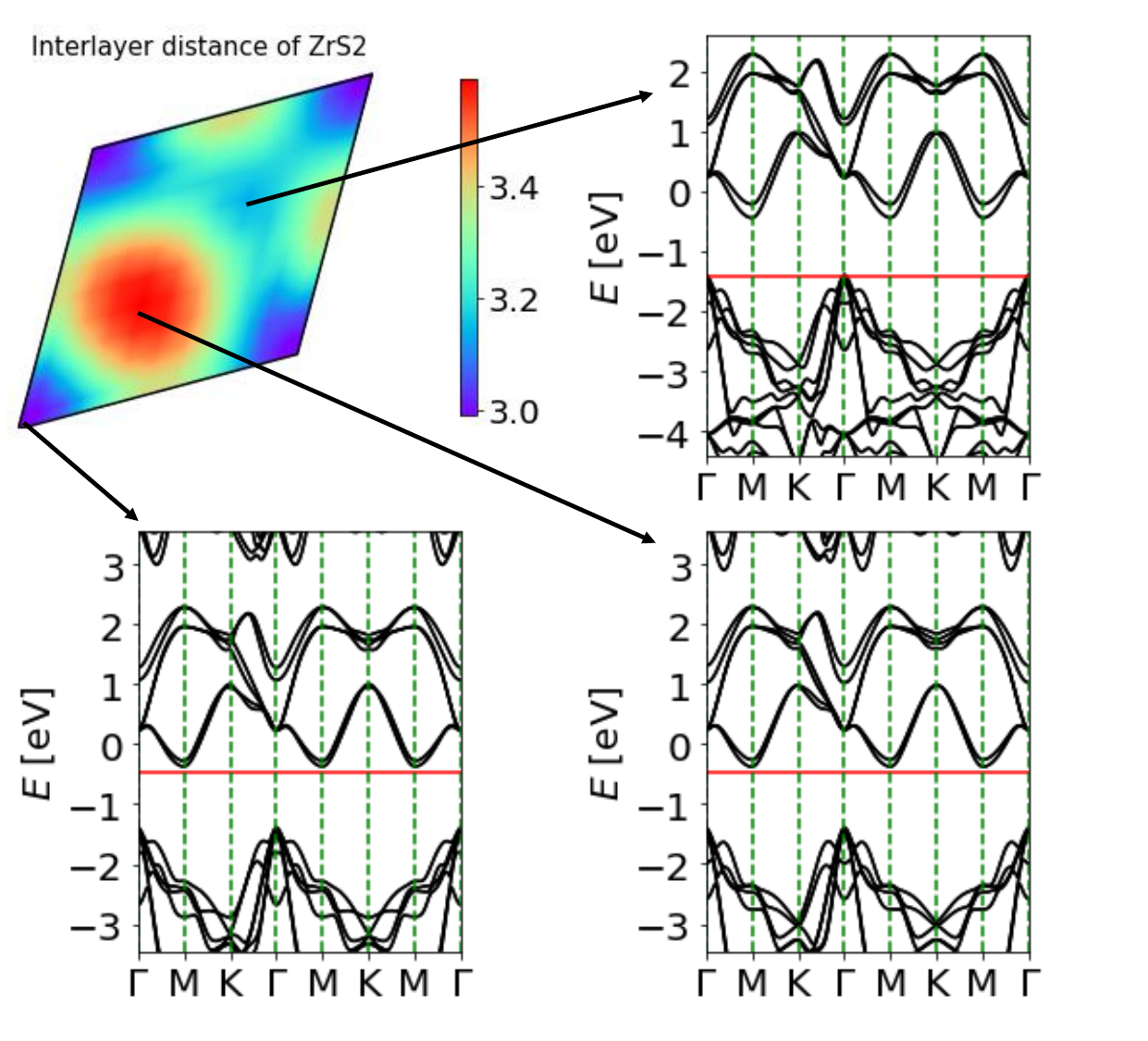}}
    \subfigure[]{\includegraphics[width=0.32\textwidth]{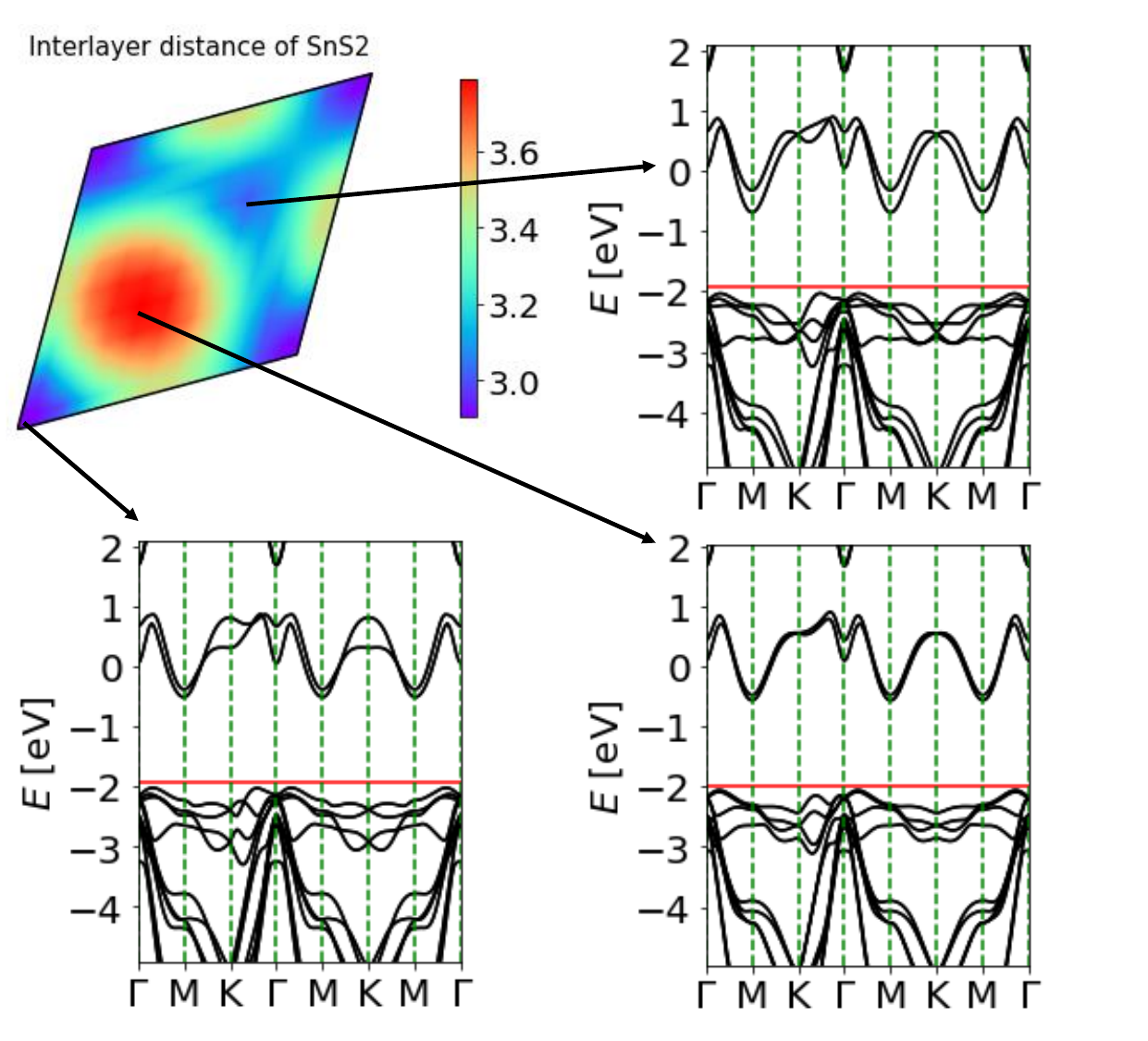}}
    \subfigure[]{\includegraphics[width=0.32\textwidth]{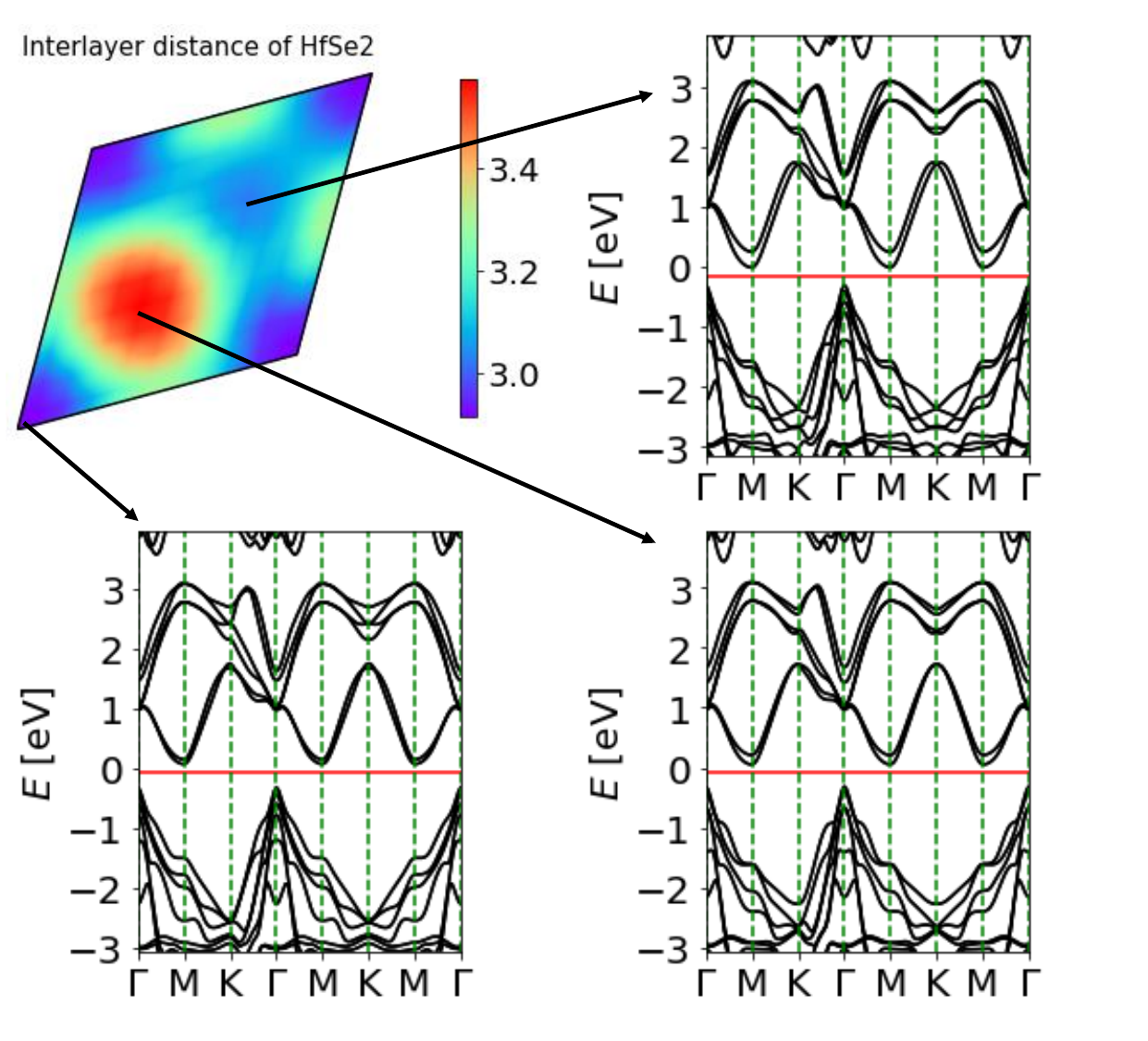}} 
    \subfigure[]{\includegraphics[width=0.32\textwidth]{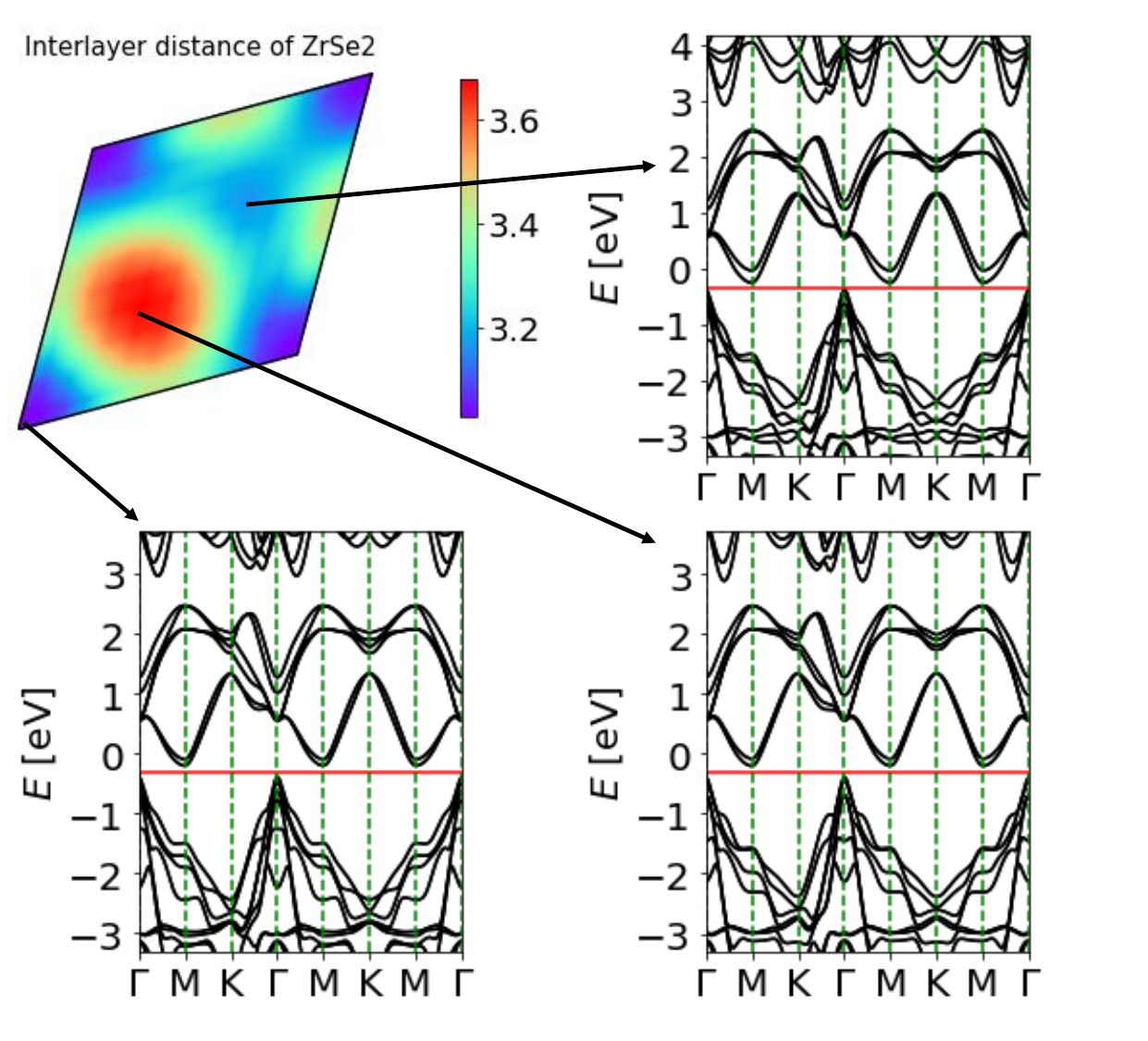}}
    \subfigure[]{\includegraphics[width=0.32\textwidth]{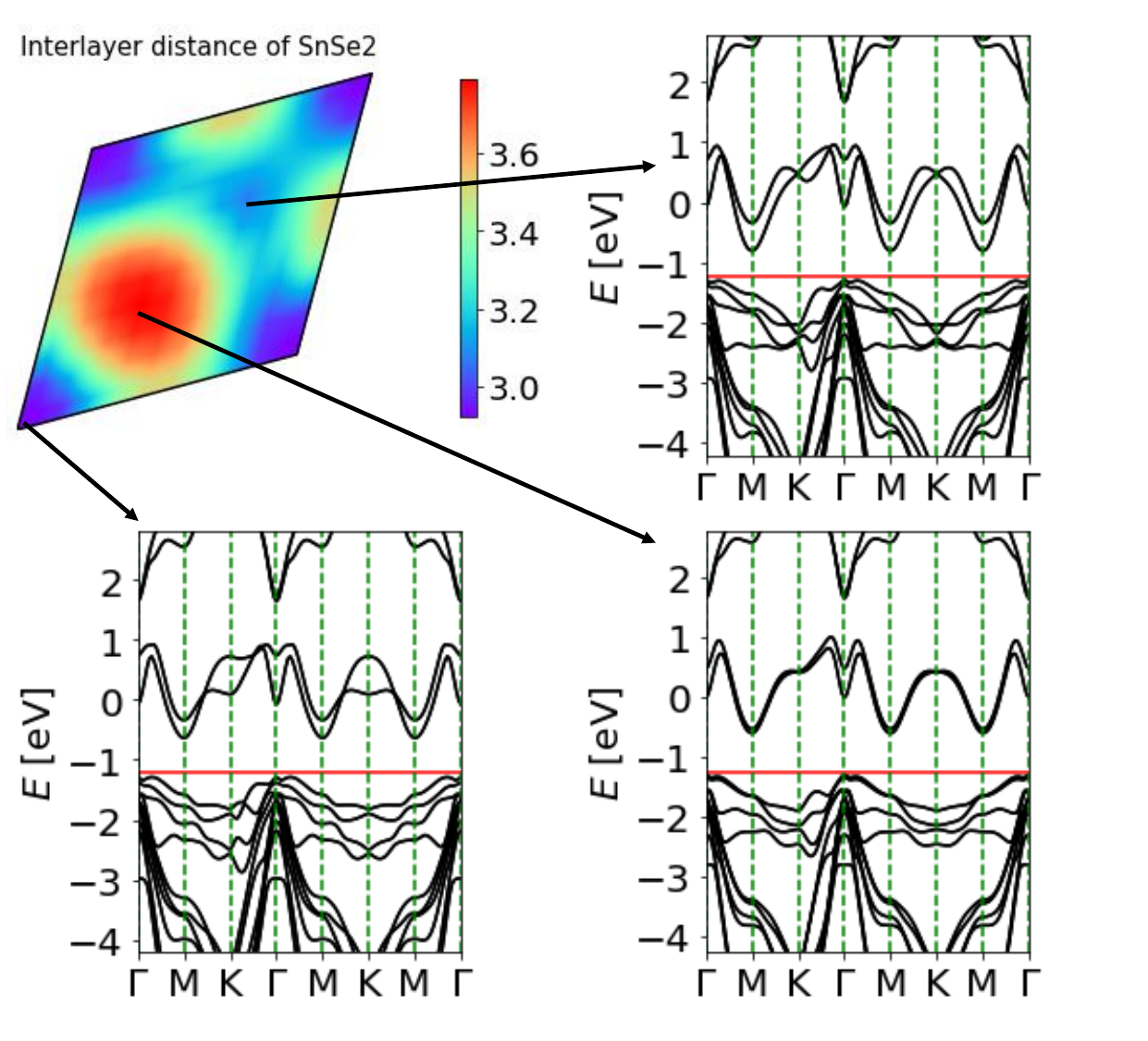}}
    \caption{Analog of Fig.~\ref{appendix:fig1} for: (a) HfS$_2$, (b) ZrS$_2$, (c) SnS$_2$, (d) HfSe$_2$, (e) ZrSe$_2$, (f) SnSe$_2$.}
    \label{fig:foobar}
\end{figure}

\begin{figure}[t!]
    \centering
    \includegraphics[width=0.49\textwidth]{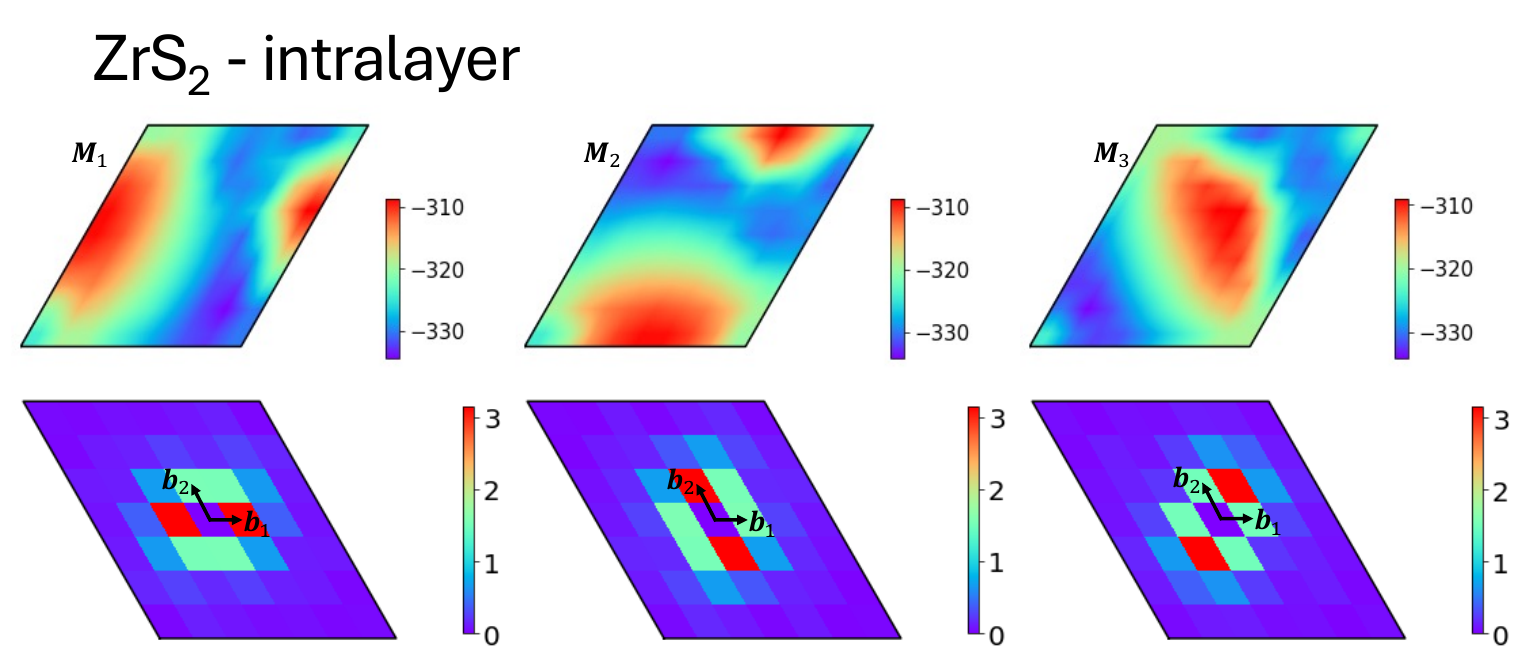}
    \includegraphics[width=0.49\textwidth]{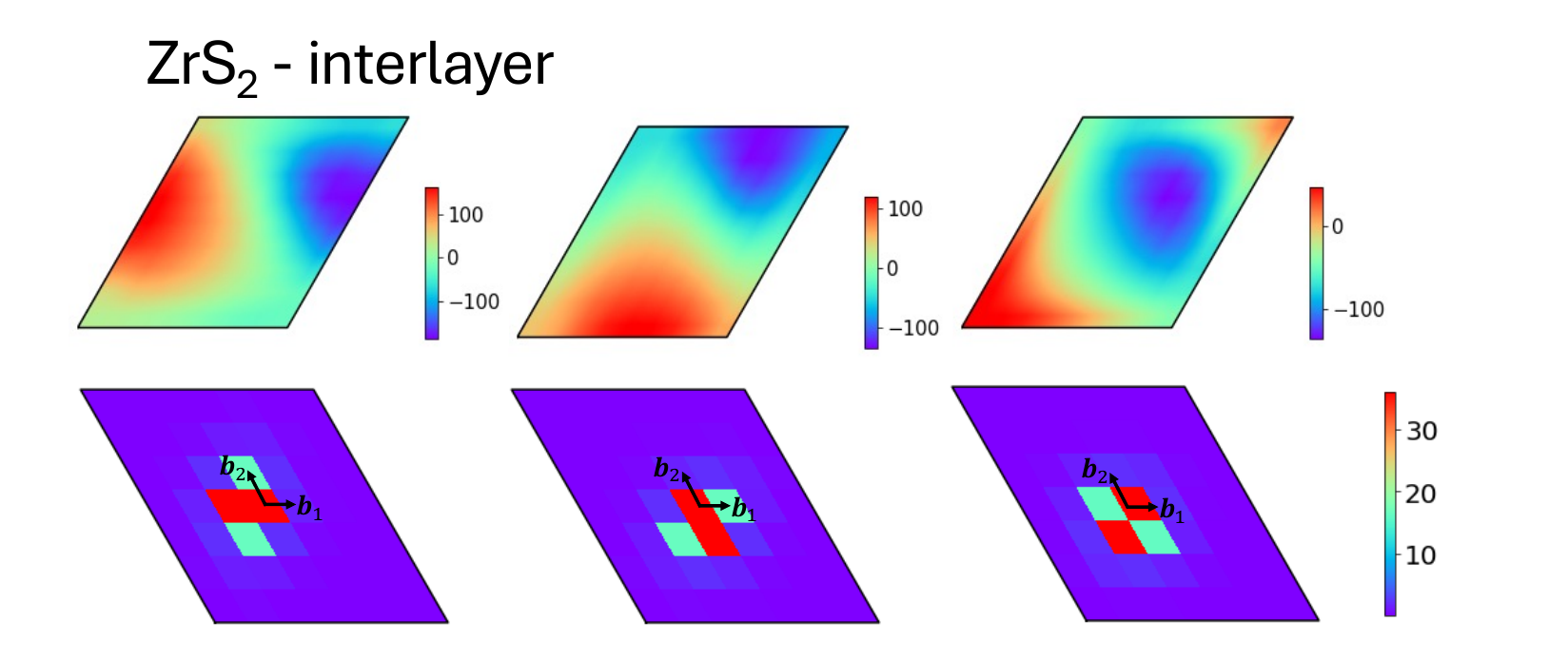}
    \includegraphics[width=0.49\textwidth]{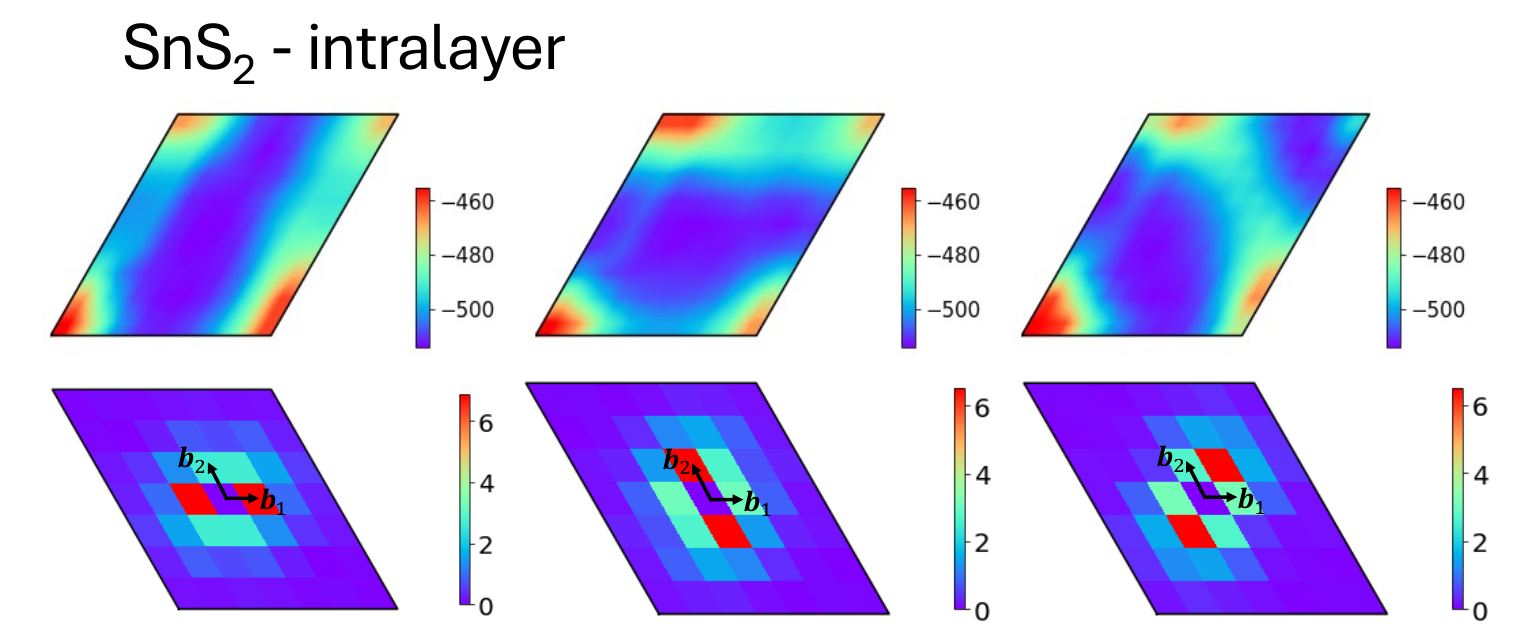}
    \includegraphics[width=0.49\textwidth]{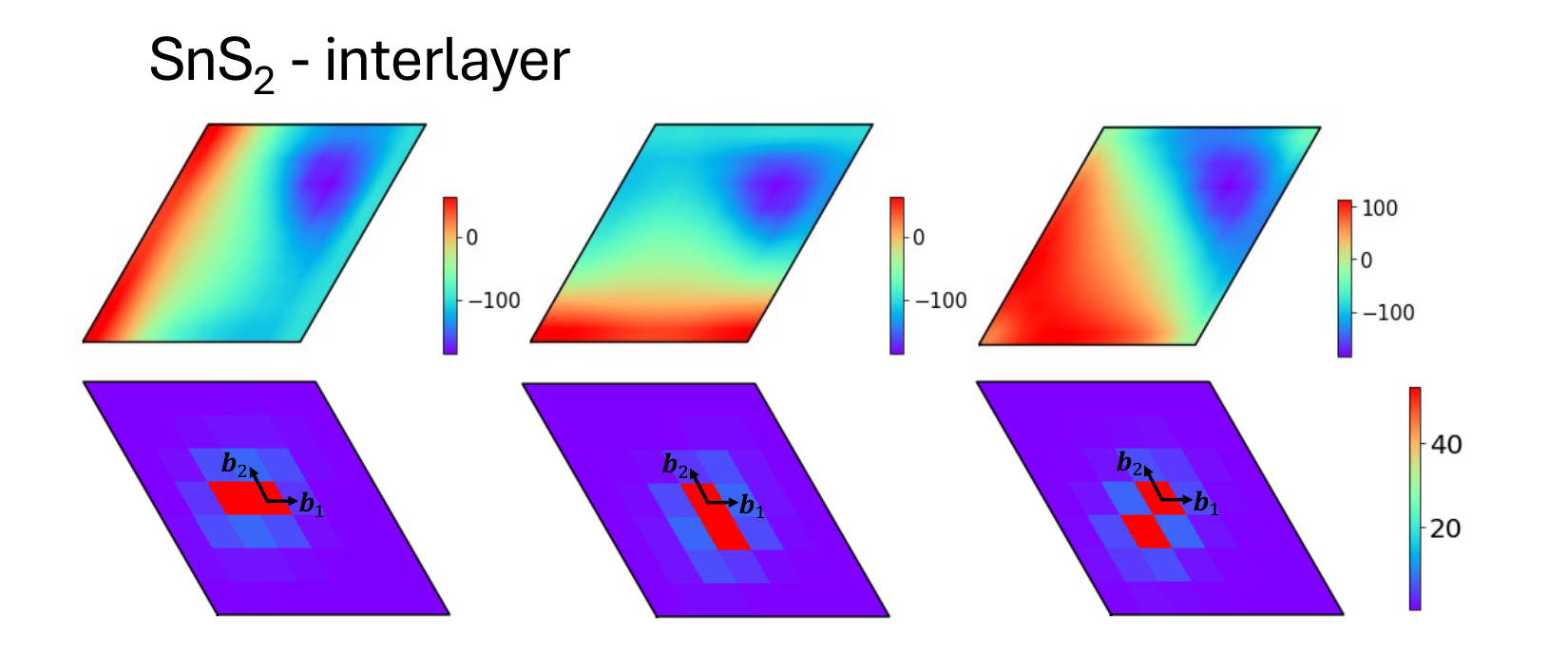}
    \caption{Analog of Figs.~\ref{fig:SI_FourierPptential} and \ref{fig:SI_FourierTunnel} for ZrS$_2$ and SnS$_2$. The Fourier component plots of MSe$_2$-based bilayers (M$=$Hf, Zr, Sn) qualitatively agree with their MS$_2$-based counterparts.}
    \label{fig:foobar2}
\end{figure}

\begin{table}[h!]
\begin{tabular}{||c c c||}
    \hline
    \, Layer type \, & \, $\bar{\Delta}_{\mathfrak{t}}^{\mathfrak{b}}(\boldsymbol{M}_{1};\boldsymbol{0})$ \, & \, $\bar{\Delta}_{\mathfrak{t}}^{\mathfrak{b}}(\boldsymbol{M}_{1};\boldsymbol{b}_{2})$ \, \\ [0.5ex] 
    \hline\hline
    HfS$_2$ & $42.01-8.0i$ & $-29.0$\\
    \hline
    HfSe$_2$ & $40.9-5.3i$ & $-24.1$ \\
    \hline\hline
    ZrS$_2$ & $35.7-6.5i$ & $-16.5$ \\
    \hline
    ZrSe$_2$ & $37.3-6.2i$ & $-16.4$  \\
    \hline\hline
    SnS$_2$ & $24.4-47.3i$ & $-7.2$  \\
    \hline
    SnSe$_2$ & $54.3-52.4i $ & $-3.1$  \\
    \hline
\end{tabular}
\caption{
Most dominant Fourier components of $\bar{\mathcal{H}}^{\boldsymbol{M}_{1}}_{\text{eff}}(\boldsymbol{k};\boldsymbol{\mathsf{d}})$ (in units of $10^{-3}$ eV, defined in the main text) 
for valley $\boldsymbol{M}_{1}$
obtained from WF-processed DFT data for rigidly displaced homobilayers of the six candidate M-valley moir\'{e} materials. 
These Fourier components should be compared against the boldfaced ones for HfS$_2$ in Table I of the main text.
The same symmetries that were described in the main text for rigidly displaced AA-stacked bilayers of 1T-HfS$_2$ are also present in homobilayers of each of the listed materials.
The effective masses and lattice constants are listed below in Table \ref{table:bandsplitting}.
\label{table:FourierComponents2}}
\end{table}

\FloatBarrier

\section{Moir\'{e} energy bands of $t$HfS$_{2}$ via the local displacement approach}
In this section we provide additional plots generated from the moir\'{e} bands model of $t$HfS$_{2}$ obtained via the local displacement approach.
\begin{figure}
	\centering
	\includegraphics[width= 0.6\linewidth]{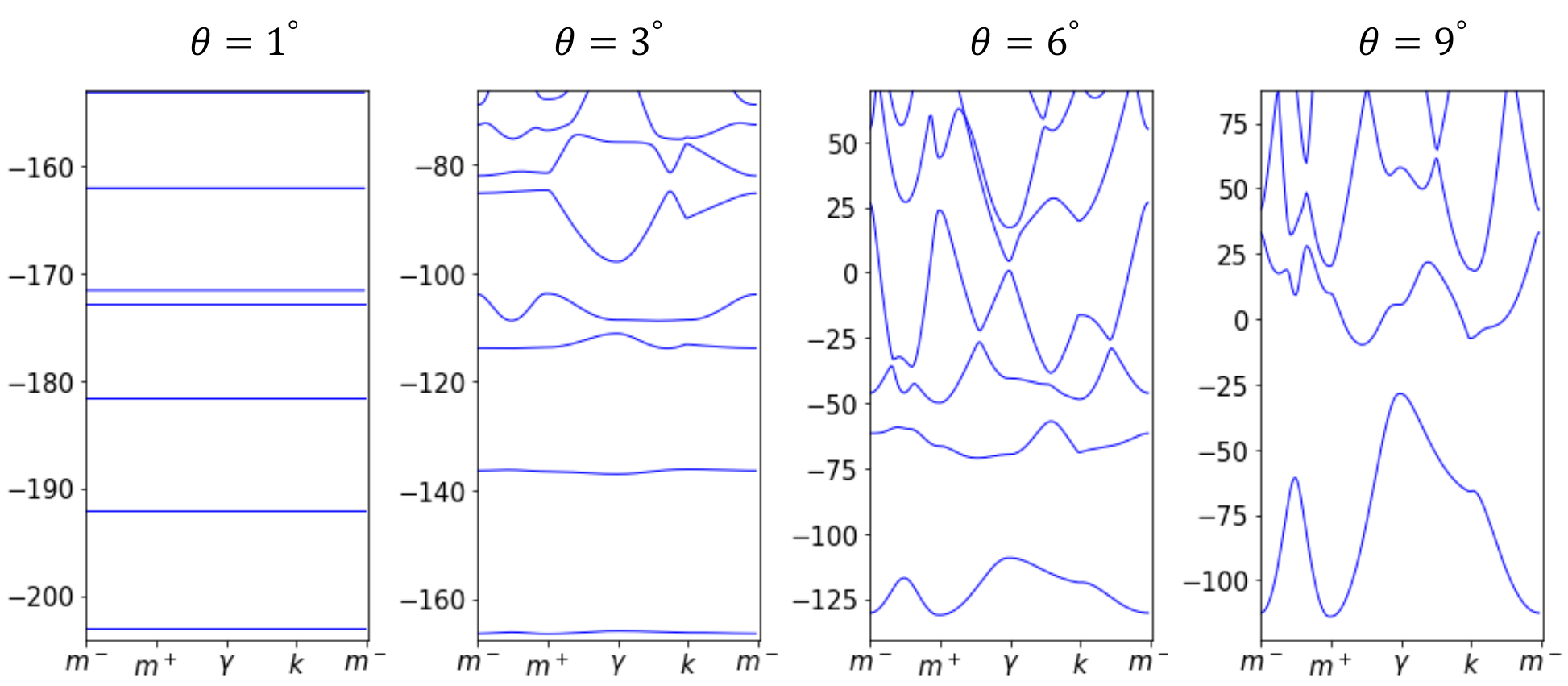}
	\caption{Bandstructure (meV) of low-energy conduction states in $t$HfS$_2$ for valley $\boldsymbol{M}_{1}$ at various twist angles. The bands in the displayed energy ranges are converged with respect to wavevector cutoff of the plane wave basis.}
	\label{fig:SI_mbands}
\end{figure}

\begin{figure}
	\centering
	\includegraphics[width= 0.6\linewidth]{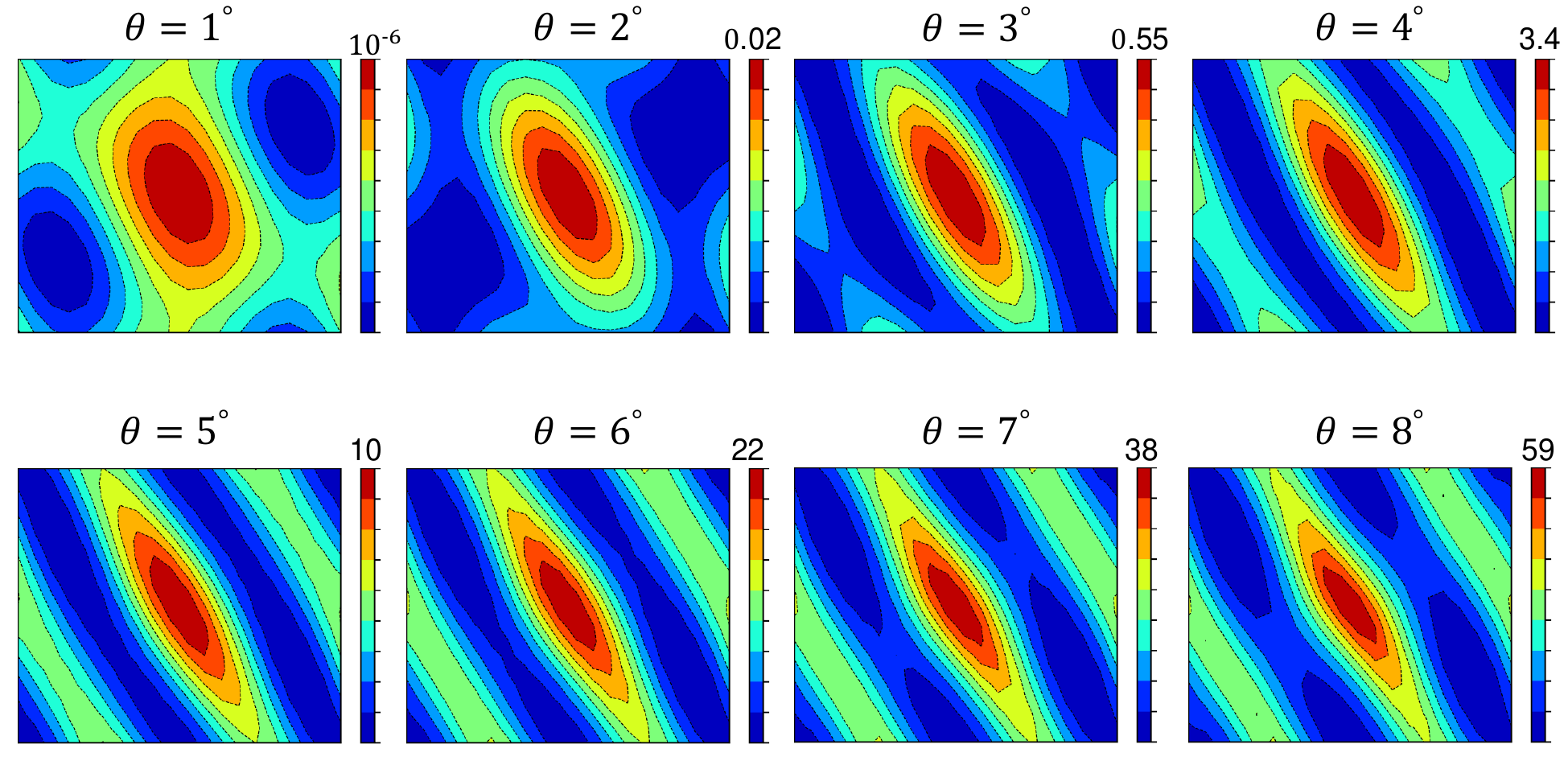}
	\caption{Contour plots of the lowest energy conduction band for valley $\boldsymbol{M}_{1}$ at various twist angles. Scale in meV and values are relative to the band minimum. Orientation is defined in Fig.~3 of the main text. In-plane relaxation cannot be neglected for twist angles below $4^{\circ}$.}
	\label{fig:SI_mcontour}
\end{figure}

To estimate the critical twist angle below which in-plane strain relaxation cannot be neglected, we use the total energy difference $\Delta V=136$ meV between low-energy and high-energy stackings (see Fig.~\ref{fig:SI_energyLandscape}), and the in-plane elastic constant of monolayer HfS$_2$ (which has isotropic elastic parameters, with diagonal value $\epsilon\approx 86$ N/m \cite{strain_hfs2}).
A characteristic length scale for domain wall width can be obtained by minimizing the summed interlayer binding energy and in-plane elastic energy. 
Assuming that similar arguments to those that apply in twisted bilayer graphene can be used in $t$HfS$_2$, 
which is not unrealistic because interlayer binding energy extrema occur at the same stacking configurations,
the characteristic width
of the domain boundary can be estimated using Eq.~(35) of Ref.~\cite{nam2017lattice}, namely
\begin{align}
    w_{d}\approx \frac{a}{4}\sqrt{\frac{2\epsilon\Omega_{uc}}{\Delta V/9}}
    \approx a\sqrt{\frac{\epsilon\Omega_{uc}}{\Delta V}}\approx 2.5 \text{ nm},
\end{align}
where $a=3.64$ \AA is the lattice constant
and $\Omega_{uc}\approx 11$ \AA$^2$ is the unit cell area of monolayer 1T-HfS$_2$. This is slightly smaller than the value $5$ nm found for bilayer graphene \cite{nam2017lattice}. Comparing $w_{d}\approx 2.5 \text{ nm}$ to the moir\'{e} lattice scale $L_{\text{M}} = a/2\sin(\theta/2)$, we find $L_{\text{M}}=w_{d}$ when $\theta\approx8^{\circ}$. 
This is the smallest angle at which domains can begin to form.
Accounting for the domain pattern in the moir\'{e} band model is important when $L_{\text{M}}\gg w_{d}$.
Thus, in-plane lattice relaxation cannot be neglected \textit{a priori} for small twist angles, say $\theta \lesssim 4^{\circ}$.\\

\section{Minimal moir\'{e} band models for candidate group IV TMDs}

A more simplistic approach to construct moir\'{e} band models is to begin from a general Hamiltonian that contains all of the symmetry allowed terms and \textit{assume} that only the first few shells of Fourier components of the $\bd$-dependent Hamiltonian matrix element are non-negligible. 
The advantage to this approach is that WFs do not need to be constructed. 
The disadvantage is that a potentially large number of parameters need to be fit and therefore the parameters that are obtained will typically be unrealistic in the absence of additional assumptions.
Since we have already explicitly demonstrated that in crystalline bilayer 1T-HfS$_2$ the first shell approximation is reliable and that certain Fourier components dominate, then, assuming crystalline homobilayers of the other candidate materials have similar physical properties, we might believe a Hamiltonian of the same form can also describe those systems. 

Consider a generic low-energy $\bk\cdot\boldsymbol{p}$ Hamiltonian for the two (doubly degenerate) lowest energy conduction bands in a rigidly displaced crystalline homobilayer of the group IV and IVB $M$-valley TMDs 1T-MX$_2$ of interest written in a basis of Bloch functions at expansion point $\bk_{0}=\boldsymbol{M}_{\nu}$ and reference displacement $\bd_{0}=\boldsymbol{0}$, one in each layer, that applies in a small patch of $\bk$-space that contains $\boldsymbol{M}_{1}=\boldsymbol{b}_{1}/2$. 
Within each spin-$s_z$ sector it is of the form
\begin{align}
	\mathcal{H}_{\bk\cdot\boldsymbol{p}}^{\boldsymbol{M}_{1}}(\boldsymbol{k};\boldsymbol{\mathsf{d}})=
	\begin{pmatrix}
		\sum_{a\in\{\parallel,\perp\}}\frac{\hbar^2}{2m_{a,l}^{\boldsymbol{M}_{1}}}\big(k^{a}-M_1^{a}\big)^2 + \Delta_{\mathfrak{b}}(\boldsymbol{M}_1;\bd) & 
        \Delta_{T}(\boldsymbol{M}_1;\bd)\\
		\Delta_{T}(\boldsymbol{M}_1;\bd)& \sum_{a\in\{\parallel,\perp\}}\frac{\hbar^2}{2m_{a,l}^{\boldsymbol{M}_{1}}}\big(k^{a}-M_1^{a}\big)^2+ \Delta_{\mathfrak{t}}(\boldsymbol{M}_1;\bd)
	\end{pmatrix},
\end{align}
where we have assumed that the anisotropic effective mass is approximately independent of $\bd$.
Notably, since each $M$ point is a time-reversal invariant momentum and since AA-stacked homobilayers have a center-of-inversion symmetry for each $\bd$, $\Delta_{T}(\boldsymbol{M}_1;\bd)$ is real-valued.
The WF-based calculations in the full model for 1T-HfS$_2$ demonstrate that it is the interlayer coupling that dominates, thus in the minimal model we neglect $\Delta_{l}(\boldsymbol{M}_1;\bd)$. 
We also found that there are four dominant Fourier components that capture the $\bd$-dependence of $\bar{\Delta}_{\mathfrak{t}}^{\mathfrak{b}}(\boldsymbol{M}_1;\bd)=e^{i\boldsymbol{M}_{1}\cdot\bd}\Delta_{T}(\boldsymbol{M}_1;\bd)$. 
That is,
\begin{align}
    \Delta_{T}(\boldsymbol{M}_1;\bd)&=e^{-i\boldsymbol{M}_{1}\cdot\bd}\bar{\Delta}_{\mathfrak{t}}^{\mathfrak{b}}(\boldsymbol{M}_1;\bd)
    =e^{-i\boldsymbol{M}_{1}\cdot\bd}\sum_{\boldsymbol{G}\in\Lambda^*}e^{-i\bG\cdot\bd}\bar{\Delta}_{\mathfrak{t}}^{\mathfrak{b}}(\boldsymbol{M}_1;\bG)\nonumber\\
    &=2\Re\left[e^{-i\boldsymbol{M}_{1}\cdot\bd}\bar{\Delta}_{\mathfrak{t}}^{\mathfrak{b}}(\boldsymbol{M}_1;\bG=\boldsymbol{0})\right]
    +2\cos((\boldsymbol{M}_1+\boldsymbol{b}_{2})\cdot\bd)\bar{\Delta}_{\mathfrak{t}}^{\mathfrak{b}}(\boldsymbol{M}_1;\boldsymbol{b}_{2})+\ldots\nonumber\\
    &\approx 2\Re\left[e^{-i\boldsymbol{M}_{1}\cdot\bd}t_{C}\right]
    +2\cos((\boldsymbol{M}_1+\boldsymbol{b}_{2})\cdot\bd)t_{R},
    \label{appendix:approx}
\end{align}
where the penultimate equality follows from the symmetry properties outlined in the main text.
Therefore, the $\bd$-dependence of the low-energy $\bk\cdot\boldsymbol{p}$ Hamiltonian for valley $\boldsymbol{M}_{1}$ is effectively encoded in two parameters:
$t_{R}\in\mathbb{R}$ and $t_{C}\in\mathbb{C}$.
(We distinguish $t_{R}$ and $t_{C}$ from $\bar{\Delta}_{\mathfrak{t}}^{\mathfrak{b}}(\boldsymbol{M}_1;\bG=\boldsymbol{0})$ and $\bar{\Delta}_{\mathfrak{t}}^{\mathfrak{b}}(\boldsymbol{M}_1;\boldsymbol{b}_{2})$ since the former will be deduced directly from DFT calculations and will serve as approximations of the latter.)
The minimal 2-orbital Hamiltonian is therefore
\begin{align}
	\mathcal{H}_{\bk\cdot\boldsymbol{p}}^{\boldsymbol{M}_{1}}(\boldsymbol{k};\boldsymbol{\mathsf{d}})=
	\begin{pmatrix}
		\sum_{a\in\{\parallel,\perp\}}\frac{\hbar^2}{2m_{a,l}^{\boldsymbol{M}_{1}}}\big(k^{a}-M_1^{a}\big)^2 & 
        T_{\boldsymbol{M}_{1}}(\bd)\\
		T_{\boldsymbol{M}_{1}}(\bd) & \sum_{a\in\{\parallel,\perp\}}\frac{\hbar^2}{2m_{a,l}^{\boldsymbol{M}_{1}}}\big(k^{a}-M_1^{a}\big)^2
	\end{pmatrix},
    \label{appendix:Hmin}
\end{align}
where $T_{\boldsymbol{M}_{1}}(\bd)\equiv 
2\Re\left[e^{-i\boldsymbol{M}_{1}\cdot\bd}t_{C}\right]+2\cos((\boldsymbol{M}_1+\boldsymbol{b}_{2})\cdot\bd)t_{R}$.

In principle, we can use the DFT data for the two lowest energy conduction bands at $\bk=\boldsymbol{M}_{1}$ for any three distinct values of $\bd$ to determine $t_{R}$ and $t_{C}$. Note, however, that each choice will result in slightly different values for $t_{R}$ and $t_{C}$ since we neglect many small but nonzero parameters in Eq.~(\ref{appendix:Hmin}).
In particular, at $\bk=\boldsymbol{M}_{1}$, we have
\begin{align}
	\mathcal{H}_{\bk\cdot\boldsymbol{p}}^{\boldsymbol{M}_{1}}(\boldsymbol{M}_{1};\boldsymbol{\mathsf{d}})=
	\begin{pmatrix}
		0 & 
        T_{\boldsymbol{M}_{1}}(\bd)\\
		T_{\boldsymbol{M}_{1}}(\bd) & 0
	\end{pmatrix}.
    \label{appendix:Hmin}
\end{align}
We choose three values of $\bd$ to consider, namely $\bd=\boldsymbol{0}$, $(\boldsymbol{a}_1+\boldsymbol{a}_2)/3$, and $2(\boldsymbol{a}_1+\boldsymbol{a}_2)/3$.
Plugging these values of $\bd$ into Eq.~(\ref{appendix:approx}), we obtain the general relations
\begin{align}
T_{\boldsymbol{M}_{1}}(\bd=\boldsymbol{0})&=2\Re\left[t_{C}\right]+2t_{R}, \nonumber\\ 
T_{\boldsymbol{M}_{1}}((\boldsymbol{a}_1+\boldsymbol{a}_2)/3)&=\Re\left[t_{C}\right]+\sqrt{3}\Im\left[t_{C}\right]-2t_{R},\nonumber\\
T_{\boldsymbol{M}_{1}}(2(\boldsymbol{a}_1+\boldsymbol{a}_2)/3)&=-\Re\left[t_{C}\right]+\sqrt{3}\Im\left[t_{C}\right]+2t_{R},
\label{appendix:rel}
\end{align}
which can be combined to yield
\begin{align}
2\sqrt{3}\Im\left[t_{C}\right] &= T_{\boldsymbol{M}_{1}}((\boldsymbol{a}_1+\boldsymbol{a}_2)/3) + T_{\boldsymbol{M}_{1}}(2(\boldsymbol{a}_1+\boldsymbol{a}_2)/3), \nonumber\\
6\Re\left[t_{C}\right] &= 2T_{\boldsymbol{M}_{1}}(\bd=\boldsymbol{0}) + T_{\boldsymbol{M}_{1}}((\boldsymbol{a}_1+\boldsymbol{a}_2)/3) - T_{\boldsymbol{M}_{1}}(2(\boldsymbol{a}_1+\boldsymbol{a}_2)/3), \nonumber\\
6t_{R} &= T_{\boldsymbol{M}_{1}}(\bd=\boldsymbol{0}) - T_{\boldsymbol{M}_{1}}((\boldsymbol{a}_1+\boldsymbol{a}_2)/3) + T_{\boldsymbol{M}_{1}}(2(\boldsymbol{a}_1+\boldsymbol{a}_2)/3).
\label{appendix:fittingEq}
\end{align}

\subsection{Hf- and Zr-based layers}
In Hf- and Zr-based layers,
at $\bd=\boldsymbol{0}$ and at $\bd=(\boldsymbol{a}_1+\boldsymbol{a}_2)/3$ we find that the lowest energy CBM state has odd layer parity (see Table \ref{table:bandsplitting}),
thus 
\begin{align}
T_{\boldsymbol{M}_{1}}(\bd=\boldsymbol{0}) > 0, \qquad
T_{\boldsymbol{M}_{1}}((\boldsymbol{a}_1+\boldsymbol{a}_2)/3)> 0.
\label{appendix:rel1}
\end{align}
At $\bd=2(\boldsymbol{a}_1+\boldsymbol{a}_2)/3$ we find that the lowest energy CBM state has even layer parity, 
thus 
\begin{align}
T_{\boldsymbol{M}_{1}}(2(\boldsymbol{a}_1+\boldsymbol{a}_2)/3)< 0.
\label{appendix:rel2}
\end{align}
Taking 1T-HfS$_2$ as an example, we find the splitting between the lowest energy conduction bands to be: $\Delta E(\bd=\boldsymbol{0})=0.049$eV, $\Delta E((\boldsymbol{a}_1+\boldsymbol{a}_2)/3)=0.172$eV, and $\Delta E(2(\boldsymbol{a}_1+\boldsymbol{a}_2)/3)=0.321$eV.
Then, using $\Delta E(\bd) = 2|T_{\boldsymbol{M}_{1}}(\bd)|$ in Eq.~(\ref{appendix:fittingEq}), as well as Eqs.~(\ref{appendix:rel1}) and (\ref{appendix:rel2}), we obtain
\begin{align}
|\Im\left[t_{C}\right]| &= |0.172-0.321|/4\sqrt{3} = 0.022, \nonumber\\
|\Re\left[t_{C}\right]| &= |2(0.049)+0.172+0.321|/12 = 0.049, \nonumber\\
|t_{R}| & = |0.049-0.172-0.321|/12 = 0.037.
\end{align}
Since $|\Re\left[t_{C}\right]| > |t_{R}|$, from the first of Eq.~(\ref{appendix:rel1}) we have $\Re\left[t_{C}\right]>0$.
Since $\Delta E(\bd=\boldsymbol{0})$ is the smallest in 1T-HfS$_2$, comparing the magnitudes of Eq.~(\ref{appendix:rel}) we see that $\Re\left[t_{C}\right]$ and $t_{R}$ must have opposite signs. Thus, $t_{R}<0$.
And since $\Delta E(2(\boldsymbol{a}_1+\boldsymbol{a}_2)/3) > \Delta E((\boldsymbol{a}_1+\boldsymbol{a}_2)/3)$, we have $\Im\left[t_{C}\right]<0$.
Examining Table \ref{table:bandsplitting}, we see that analogous arguments apply in all of the Hf- and Zr-based TMDs that we consider. The results are listed in Table \ref{table:k.p}.

\subsection{Sn-based layers}
In Sn-based layers,
at $\bd=\boldsymbol{0}$ we find that the lowest energy CBM state has odd layer parity, 
thus 
\begin{align}
T_{\boldsymbol{M}_{1}}(\bd=\boldsymbol{0})> 0.
\label{appendix:rel3}
\end{align}
At $\bd=(\boldsymbol{a}_1+\boldsymbol{a}_2)/3$ and at $\bd=2(\boldsymbol{a}_1+\boldsymbol{a}_2)/3$ we find that the lowest energy CBM state has even layer parity, 
thus 
\begin{align}
T_{\boldsymbol{M}_{1}}((\boldsymbol{a}_1+\boldsymbol{a}_2)/3) < 0, \qquad
T_{\boldsymbol{M}_{1}}(2(\boldsymbol{a}_1+\boldsymbol{a}_2)/3) < 0.
\label{appendix:rel4}
\end{align}
Taking 1T-SnS$_2$ as an example, we find the splitting between the lowest energy conduction bands to be: $\Delta E(\bd=\boldsymbol{0})=0.134$eV, $\Delta E((\boldsymbol{a}_1+\boldsymbol{a}_2)/3)=0.093$eV, and $\Delta E(2(\boldsymbol{a}_1+\boldsymbol{a}_2)/3)=0.365$eV.
Then, using $\Delta E(\bd) = 2|T_{\boldsymbol{M}_{1}}(\bd)|$ in Eq.~(\ref{appendix:fittingEq}), as well as Eqs.~(\ref{appendix:rel3}) and (\ref{appendix:rel4}), we obtain
\begin{align}
|\Im\left[t_{C}\right]| &= |0.093+0.365|/4\sqrt{3} = 0.066, \nonumber\\
|\Re\left[t_{C}\right]| &= |2(0.134)-0.093+0.365|/12 = 0.045, \nonumber\\
|t_{R}| & = |0.134+0.093-0.365|/12 = 0.012.
\end{align}
Since $T_{\boldsymbol{M}_{1}}((\boldsymbol{a}_1+\boldsymbol{a}_2)/3) < 0$ and
$T_{\boldsymbol{M}_{1}}(2(\boldsymbol{a}_1+\boldsymbol{a}_2)/3)<0$, from Eq.~(\ref{appendix:rel})
we have that $\Im\left[t_{C}\right]<0$.
And since $\Delta E(2(\boldsymbol{a}_1+\boldsymbol{a}_2)/3) > \Delta E((\boldsymbol{a}_1+\boldsymbol{a}_2)/3)$ we also have $\Re\left[t_{C}\right]-2t_{R} > 0$, thus $\text{Re}[t_{C}]>0$. 
We can now simply plug in the parameter values with known signs and deduce the sign of $t_{R}$ by comparing to the known value of $\Delta E(\bd)$ for one value of $\bd$.
Then, for 1T-SnS$_2$ we have $0.134=\Delta E(\bd=\boldsymbol{0})=2|T_{\boldsymbol{M}_{1}}(\bd=\boldsymbol{0})|=2*45\pm2|t_{R}|$, thus $t_{R}=-12$.

\begin{table}[t!]
\begin{tabular}{||c c c c||}
    \hline
    \, Layer type \, & \, $\Delta E(\bd=\boldsymbol{0})$ eV \, & \, $\Delta E((\boldsymbol{a}_1+\boldsymbol{a}_2)/3)$ eV \, & \, $\Delta E(2(\boldsymbol{a}_1+\boldsymbol{a}_2)/3)$ eV \\ [0.5ex] 
    \hline\hline
    HfS$_2$ & 0.049 (odd) & 0.172 (odd) & 0.321 (even)  \\
    \hline
    HfSe$_2$ & 0.079 (odd) & 0.157 (odd) & 0.263 (even) \\
    \hline\hline
    ZrS$_2$ & 0.095 (odd) & 0.119 (odd) & 0.227 (even) \\
    \hline
    ZrSe$_2$ & 0.108 (odd) & 0.121 (odd) & 0.220 (even) \\
    \hline\hline
    SnS$_2$ & 0.134 (odd) & 0.093 (even) & 0.365 (even) \\
    \hline
    SnSe$_2$ & 0.297 (odd) & 0.071 (even) & 0.461 (even) \\
    \hline
\end{tabular}
\caption{Energy difference between lowest energy (doubly degenerate) conduction bands and parity of the lowest energy conduction state at $\boldsymbol{M}_{1}$ for three values of relative in-plane layer displacement $\bd$.
\label{table:bandsplitting}}
\end{table}

\begin{table}[t!]
\begin{tabular}{||c c c c c||}
    \hline
    \, Layer type \, & \, $t_{C}$ (meV) \, & \, $t_{R}$ (meV) \, & \, $m_{\parallel,l}^{\boldsymbol{M}_{1}}$ ($m_e$) \, & \, $m_{\perp,l}^{\boldsymbol{M}_{1}}$ ($m_e$) \,\\ [0.5ex] 
    \hline\hline
    HfS$_2$ & $49 - 22 i$ & $-37$& $2.41$ & $0.27$ \\
    \hline
    HfSe$_2$ & $ 48 - 15 i$ & $- 28$ & $2.11$ &  $0.20$ \\
    \hline\hline
    ZrS$_2$ & $ 45 - 16 i$ & $- 21$ & $1.89$ & $0.29$ \\
    \hline
    ZrSe$_2$ & $ 46 - 14 i$ & $- 19$ & $1.81$ & $0.21$ \\
    \hline\hline
    SnS$_2$ & $45 - 66 i$ & $- 12$ &  $1.00$ & $0.25$ \\
    \hline
    SnSe$_2$ & $82 - 77 i$ & $- 8$ & $0.71$ & $0.21$ \\
    \hline
\end{tabular}
\caption{Parameters (rounded to $1$ meV) for a minimal description of the two lowest energy conduction band states near the $\boldsymbol{M}_1$ valley in candidate $M$ valley 1T-MX$_2$ TMD moir\'{e} homobilayers, assuming that the same parameters that dominate for HfS$_2$ also dominate in the other materials. This assumption is supported by the more complete WF-based approach described above.
We also list the effective masses $m_{\parallel,l}^{\boldsymbol{M}_{1}}$ and $m_{\perp,l}^{\boldsymbol{M}_{1}}$ in units of the bare electron mass $m_e$.}
\label{table:k.p}
\end{table}


\bibliography{moireHfS2}